\documentclass[aps,prb,twocolumn,nobibnotes]{revtex4-1}

\usepackage[pdftex]{graphicx}
\usepackage{amssymb,amsmath}
\usepackage{nicefrac}
\usepackage{physics}
\usepackage{color}
\usepackage{braket}
\usepackage{mathptmx}
\usepackage[T1]{fontenc}
\usepackage{bm}
\usepackage{subfloat}
\usepackage{soul}
\DeclareSymbolFont{boldoperators}{OT1}{cmr}{bx}{n}
\SetSymbolFont{boldoperators}{bold}{OT1}{cmr}{bx}{n}
\edef\bar{\unexpanded{\protect\mathaccentV{bar}}\number\symboldoperators16}
\begin{document}

\def\bra#1{\left<{#1}\right|}
\def\ket#1{\left|{#1}\right>}
\def\expval#1#2{\bra{#2} {#1} \ket{#2}}
\def\mapright#1{\smash{\mathop{\longrightarrow}\limits^{_{_{\phantom{X}}}{#1}_{_{\phantom{X}}}}}}

\title{On the calculation of quantum mechanical electron transfer rates}
\author{Joseph E. Lawrence}
\email[Email: ]{joseph.lawrence@chem.ox.ac.uk}
\affiliation{Department of Chemistry, University of Oxford, Physical and Theoretical Chemistry Laboratory, South Parks Road, Oxford, OX1 3QZ, UK}
\author{Theo Fletcher}
\affiliation{Department of Chemistry, University of Oxford, Physical and Theoretical Chemistry Laboratory, South Parks Road, Oxford, OX1 3QZ, UK}
\author{Lachlan P. Lindoy}
\affiliation{Department of Chemistry, University of Oxford, Physical and Theoretical Chemistry Laboratory, South Parks Road, Oxford, OX1 3QZ, UK}
\author{David E. Manolopoulos}
\affiliation{Department of Chemistry, University of Oxford, Physical and Theoretical Chemistry Laboratory, South Parks Road, Oxford, OX1 3QZ, UK}

\begin{abstract}
We present a simple interpolation formula for the rate of an electron transfer reaction as a function of the electronic coupling strength. The formula only requires the calculation of Fermi Golden Rule and Born-Oppenheimer rates and so can be combined with any methods that are able to calculate these rates. We first demonstrate the accuracy of the formula by applying it to a one dimensional scattering problem for which the exact quantum mechanical, Fermi Golden Rule, and Born-Oppenheimer rates are readily calculated. We then describe how the formula can be combined with the Wolynes theory approximation to the Golden Rule rate, and the ring polymer molecular dynamics (RPMD) approximation to the Born-Oppenheimer rate, and used to capture the effects of nuclear tunnelling, zero point energy, and solvent friction on condensed phase electron transfer reactions. Comparison with exact hierarchical equations of motion (HEOM) results for a demanding set of spin-boson models shows that the interpolation formula has an error comparable to that of RPMD rate theory in the adiabatic limit, and that of Wolynes theory in non-adiabatic limit, and is therefore as accurate as any method could possibly be that attempts to generalise these methods to arbitrary electronic coupling strengths. \end{abstract}

\maketitle

\section{Introduction}
There has been a steady development of imaginary time path integral methods for including the effects of nuclear tunnelling and zero point energy on the rates of electronically {\em adiabatic} chemical reactions. The earliest was the thermal instanton approximation introduced by Miller in 1975, which uses a single imaginary time trajectory to capture the effect of nuclear tunnelling.\cite{Miller75}  Although this approximation can be very accurate, it is only applicable to problems in which the rate is dominated by a single tunnelling path.\cite{Richardson09,Andersson09,Richardson18} It cannot be used for reactions in solution, in which the configuration space of the solvent leads to multiple contributing tunnelling paths. In order to overcome this problem, methods which use statistical sampling of imaginary time paths are required. The earliest such methods were quantum mechanical generalisations of transition state theory, such as centroid density quantum transition state theory (QTST)\cite{Voth89} and the quantum instanton model.\cite{Miller03,Vanicek05} While these methods can be applied to reactions in solution, they rely on the specification of a transition state dividing surface, a poor choice of which can lead to a significant overestimation of the  rate constant. This final difficulty was overcome with the introduction of ring polymer molecular dynamics\cite{Craig04} (RPMD) reaction rate theory,\cite{Craig05a,Craig05b} which contains a dynamical correction to QTST and gives a rate that is rigorously independent of the choice of dividing surface. There is now ample evidence that RPMD rate theory has effectively solved the problem of estimating quantum mechanical reaction rates in complex systems,\cite{Collepardo08,Boekelheide11,Habershon13} provided the Born-Oppenheimer approximation is valid. 

For reactions that involve transitions from one electronic state to another, such as electron transfer and proton-coupled electron transfer reactions, the Born-Oppenheimer approximation cannot be made. The most widely used method for understanding these reactions is Marcus theory,\cite{Marcus85} which assumes the reaction is in the {\em non-adiabatic} (Fermi Golden Rule) limit and that the nuclei can be treated classically with harmonic free energy surfaces. As in the adiabatic case, one can go beyond the classical Marcus treatment by using instanton approaches\cite{Richardson15a,Richardson15b} and path integral sampling methods\cite{Wolynes87,Lawrence18} to include quantum mechanical effects in the nuclear motion. And again as in the adiabatic case, one needs to use a path integral sampling method to compute the rate in a complex system such as a liquid, in which there can be many contributing tunnelling paths. The earliest such method was Wolynes theory.\cite{Wolynes87} This is based on a steepest descent approximation to the time integral of a reactive flux autocorrelation function in the non-adiabatic limit, which leads to a simple imaginary time path integral expression for the rate constant that can readily be applied to electron transfer reactions in complex systems.\cite{Wolynes89,Bader90,Wolynes91}

Appropriate computational methods therefore exist for calculating chemical reaction rates in both the adiabatic and non-adiabatic limits. However, many real systems are not fully in either of these limits. Sometimes this is because the separation between the adiabats is neither perturbatively small, nor is it large enough that the Born-Oppenheimer approximation is valid. In the case of electron transfer, it is well known that solvent friction can drive a reaction away from the non-adiabatic limit even when the electronic coupling is very small.\cite{Zusman80,Garg85,Hynes86,Rips87,Sparpaglione88} There exist several analytical theories that are able to describe the transition from non-adiabatic to adiabatic behaviour. For example, Zusman theory\cite{Zusman80} extends Marcus theory\cite{Marcus85} to systems with strong solvent friction. However, there currently exists no generally applicable and reliable simulation method that can be used to calculate electron transfer rates for arbitrarily complex systems between the adiabatic and non-adiabatic limits, and that also includes the effects of zero point energy and tunnelling. 

Because of the success of RPMD in the adiabatic limit, there have been several attempts to generalise it to treat non-adiabatic systems.\cite{Shushkov12,Richardson13,Ananth13,Duke16,Menzeleev14,Kretchmer16,Chowdhury17,Tao18,Thapa19} However each of these has its issues. For example, kinetically constrained RPMD,\cite{Menzeleev14,Kretchmer16} which introduces a \color{black} dynamical \color{black} electronic variable and a corresponding \color{black}system-dependent \color{black} electronic mass parameter, \color{black}has been shown to be quantitatively inaccurate in the non-adiabatic limit for a series of relatively simple spin-boson models.\cite{Kretchmer16} (Despite this, it has been successfully applied to a study of electron transfer in a protein environment, where it was used to reveal an unexpected reaction mechanism.\cite{Kretchmer18}) \color{black} Non-adiabatic RPMD,\cite{Richardson13} which is based on the well-known (Meyer-Miller,\cite{Meyer79} Stock-Thoss\cite{Stock97}) classical electron mapping approach, does not correctly preserve the quantum mechanical Boltzmann distribution. Mapping variable RPMD\cite{Ananth13} is a modified version of non-adiabatic RPMD which {\em does} preserve the quantum Boltzmann distribution, but it suffers from a pathological sign problem that arises from a cancellation between positively and negatively weighted regions of the ring polymer phase space. \color{black} Standard Born-Oppenheimer RPMD rate theory can be applied to some electron transfer reactions by explicitly including an electron as a distinguishable particle, but this is limited to systems for which using a one electron pseudo-potential is a valid approximation.\cite{Menzeleev11,Kretchmer13} \color{black} Similar objections can be raised to all of the other methods that have been suggested so far to extend RPMD to electronically non-adiabatic reactions,\cite{Shushkov12,Richardson13,Ananth13,Duke16,Menzeleev14,Kretchmer16,Chowdhury17,Tao18} none of which is entirely satisfactory.

Here we suggest a simple method for including nuclear quantum effects in \color{black} the calculation of electron transfer rates \color{black} between the adiabatic and non-adiabatic limits that avoids the need to find a non-adiabatic generalisation of RPMD. In particular, we show how one can perform separate Golden Rule and Born-Oppenheimer simulations and then interpolate between their results to calculate the electron transfer rate for an arbitrary electronic coupling strength. Since there already exist well established methods for calculating the Born-Oppenheimer and Golden Rule rates, the method we are suggesting is immediately applicable to realistic simulations of condensed phase electron transfer reactions. Furthermore, we shall show that using the interpolation formula gives an error comparable to that of RPMD rate theory in the adiabatic limit, and Wolynes theory in the non-adiabatic limit, and is therefore as good as one could possibly hope to do by generalising RPMD to treat non-adiabatic reactions. \color{black} Our interpolation formula is closely related to earlier expressions for the rate in high-temperature limit of the spin-boson model,\cite{Zusman80,Garg85,Hynes86,Rips87,Sparpaglione88,Calef83,Pollak90,Rips95,Starobinets96,Gladkikh05} in particular to expressions based on Pad\'e resummations of the perturbative series in the electronic coupling strength. And as we shall show in Section~\ref{interpol_sec}, it reduces to a known analytical result, the Zusman equation,\cite{Zusman80} in the appropriate limit.\color{black}

Section~\ref{Theory_Section} summarises electron transfer rate theory and describes how Wolynes theory can be used to calculate the Golden Rule rate, and RPMD reaction rate theory the Born-Oppenheimer rate, for a typical condensed phase electron transfer reaction. Section~\ref{interpol_sec} introduces our new formula for interpolating between the results of these calculations to obtain an expression for the electron transfer rate at intermediate electronic coupling strengths, and highlights some of its properties. Section~\ref{Scattering_Section} investigates the accuracy of the interpolation formula for a simple one-dimensional scattering problem for which the exact quantum mechanical, Golden Rule, and Born-Oppenheimer rates are straightforward to calculate. Section~\ref{spin_b_sec} combines the formula with the RPMD approximation to the Born-Oppenheimer rate and the Wolynes theory approximation to the Golden Rule rate and applies it to the spin-boson model in a demanding set of regimes -- with small and large nuclear quantum effects, low and high electronic coupling strengths, weak and strong solvent frictions, and two very different thermodynamic driving forces. The results are shown to agree well with exact quantum mechanical rate constants obtained using the hierarchical equations of motion (HEOM) method\cite{Tanimura89,Tanimura90,Tanimura06,Yan04,Xu05,Xu07,Jin08} in all of these regimes. Section~VI concludes the paper, leaving an appendix to describe an efficient implementation of RPMD for system-bath problems such as the spin-boson model considered in Section~V.
%\vfill\null

\section{Theoretical Background} \label{Theory_Section}

\subsection{Electron transfer rates}

The Hamiltonian for a condensed phase electron transfer reaction can be written in the diabatic representation as
\begin{equation}
\hat{H}=\hat{H}_0\dyad{0}{0}+\hat{H}_1\dyad{1}{1} + \Delta(\dyad{0}{1}+\dyad{1}{0}),
\end{equation}
where
\begin{equation}
\hat{H}_i = \sum_{\nu=1}^{f} \frac{\hat{p}^2_{\nu}}{2m_{\nu}} + {V}_i(\hat{\bm{q}}).
\end{equation}
Here $\hat{H}_i$ is the nuclear Hamiltonian on state $i$ with diabatic potential ${V}_i(\bm{q})$, and $\Delta$  is the diabatic electronic coupling, which we shall take here to be a constant (the Condon approximation).\cite{Nitzan06}

The exact quantum mechanical rate of transfer from state $\ket{0}$ to state $\ket{1}$ can be obtained by considering how the populations of the two states return to equilibrium when the system is prepared on state $\ket{0}$ with the initial density operator
\begin{equation}
\hat{\rho}(0) = {1\over\beta Q_r}\int_0^{\beta}{\rm d}\lambda\,e^{-(\beta-\lambda)\hat{H}}\hat{P}_r\, e^{-\lambda\hat{H}},
\end{equation}
where $\hat{P}_r=\dyad{0}{0}$ is the projection operator onto the reactant state, $Q_r$ is the reactant partition function
\begin{equation}
Q_r = {\rm Tr}\left[e^{-\beta\hat{H}}\hat{P}_r\right],
\end{equation}
and $\beta=1/k_{\rm B}T$. Provided the dynamics of the electron transfer is such that a kinetic description is valid, the time-dependent populations of the two states 
\begin{equation}
P_{s}(t) = {\rm Tr}\left[\hat{\rho}(0)\hat{P}_s(t)\right],
\end{equation}
where
\begin{equation}
\hat{P}_{s}(t) = e^{+i\hat{H}t/\hbar}\hat{P}_{s}\,e^{-i\hat{H}t/\hbar}
\end{equation}
with $s=r$ and $p$ and $\hat{P}_p=\dyad{1}{1}$, will eventually settle down to satisfy the kinetic equations
\begin{equation}
\dot{P}_r(t) = -k_{}P_r(t)+k'P_p(t),
\end{equation}
\begin{equation}
\dot{P}_p(t) = -k'P_p(t)+k_{}P_r(t),
\end{equation}
in which $k_{}$ and $k'$ are the forward and backward electron transfer rate constants. Using the fact that $P_r(t)+P_p(t)=1$ in these equations gives
\begin{equation}
\dot{P}_p(t) = k_{}-(k_{}+k')P_p(t) = (k_{}+k')(\left<P_p\right>-P_p(t))
\end{equation}
and therefore\cite{Craig07}
\begin{equation}
k_{} = \lim_{t\to\infty} {\dot{P}_p(t)\over 1-P_p(t)/\left<P_p\right>},
\end{equation}
where (by detailed balance)
\begin{equation}
\left<P_p\right> = \lim_{t\to\infty} P_p(t) = {{\rm Tr}\left[e^{-\beta\hat{H}}\hat{P}_p\right]\over {\rm Tr}\left[e^{-\beta\hat{H}}\right]}
\end{equation}
is the thermal equilibrium population of the product state and we have introduced the limit as ${t\to\infty}$ in Eq.~(10) to stress that the kinetic description only becomes appropriate after an initial transient period. (As $t\to\infty$, both the numerator and denominator of Eq.~(10) tend to zero, but their ratio remains finite and tends to $k_{}$.)

Combining Eqs.~(3), (5), and (6), one finds that the numerator of Eq.~(10) can be re-written as
\begin{equation}
\dot{P}_p(t) = {\tilde{c}_{fs}(t)\over Q_r}
\end{equation}
where
\begin{equation}
\tilde{c}_{fs}(t) = {1\over\beta}\int_0^{\beta} {\rm d}\lambda\,{\rm Tr}\left[e^{-(\beta-\lambda)\hat{H}}\hat{F}_p\,e^{-\lambda\hat{H}}\hat{P}_p(t)\right]
\end{equation}
is a Kubo-transformed\cite{Kubo57} flux-side correlation function, with
\begin{equation}
\hat{F}_p=-{i\over\hbar}\left[\hat{H},\hat{P}_r\right] = +{i\over\hbar}\left[\hat{H},\hat{P}_p\right] 
= {i\Delta\over\hbar}\left(\dyad{0}{1}-\dyad{1}{0}\right).
\end{equation}
This makes a connection with the original linear response theory derivation of the forward quantum mechanical rate constant by Yamamoto,\cite{Yamamoto60} who argued that Eq.~(10) can often be replaced to a good approximation by
\begin{equation}
k_{} \simeq {\tilde{c}_{fs}(t_{\rm p})\over Q_r}
\end{equation}
for some appropriate \lq\lq plateau time'' $t_{\rm p}$. This plateau time is supposed to be sufficiently short that very little population has been transferred to the product state ($P_p(t_{\rm p})\ll \left<P_p\right>$), and yet sufficiently long that the population dynamics has settled down to conform to the kinetic behaviour in Eqs.~(7) and~(8). Clearly, therefore, Eq.~(15) rests on a separation of time scales, and it will be most accurate for slow electron transfer reactions. In fact, Eq.~(15) becomes {\em exact} in the non-adiabatic ($\Delta\to 0$) limit that we shall discuss next, even as $t_{\rm p}\to \infty$ (the two limits being taken in such a way that $\Delta^2 t_{\rm p}\to 0$). However, away from the non-adiabatic limit, Eq.~(15) only provides an approximation to the electron transfer rate, whereas Eq.~(10) remains exact. We shall therefore return to Eq.~(10) in Sec.~II.C.

\subsection{Wolynes theory}

As we have just discussed, the forward electron transfer rate in the non-adiabatic ($\Delta\to 0$) limit can be calculated as
\begin{equation}
k_{} = \lim_{t_{\rm p}\to\infty} {\tilde{c}_{fs}(t_{\rm p})\over Q_r},
\end{equation}
or equivalently as
\begin{equation}
k_{} = {1\over 2Q_r}\int_{-\infty}^{\infty} \tilde{c}_{ff}(t)\,{\rm d}t,
\end{equation}
where 
\begin{equation}
\tilde{c}_{ff}(t) = {1\over\beta}\int_0^{\beta} {\rm d}\lambda\,{\rm Tr}\left[e^{-(\beta-\lambda)\hat{H}}\hat{F}_p\,e^{-\lambda\hat{H}}\hat{F}_p(t)\right]
\end{equation}
is a flux-flux correlation function.\cite{Miller83} In fact, the time integral (from $-\infty$ to $\infty$) of the trace in this expression can be shown to be independent of $\lambda$, and so we can equally well write
\begin{equation}
k_{} = {1\over 2Q_r}\int_{-\infty}^{\infty} {\rm Tr}\left[e^{-(\beta-\lambda)\hat{H}}\hat{F}_p\,e^{-\lambda\hat{H}}\hat{F}_p(t)\right]\,{\rm d}t,
\end{equation}
for any value of $\lambda$ in the range $0\le\lambda\le\beta$. If we now use the expression for $\hat{F}_p$ in Eq.~(14), and retain just the two leading contributions of $O(\Delta^2)$, we find (after a little algebra) that this reduces to the Fermi Golden Rule approximation to the rate constant\cite{Wolynes87,Sparpaglione88}
\begin{equation}
k_{\rm GR}\!=\! \frac{\Delta^2}{\hbar^2 Q_r}\! \int_{-\infty}^{\infty}\! \tr\left[e^{-(\beta-\lambda)\hat{H}_0} e^{-i \hat{H}_0t/\hbar}e^{-\lambda\hat{H}_1}e^{+i \hat{H}_1t/\hbar}\right] \mathrm{d}t, \label{FGR_rate}
\end{equation}
in which the reactant partition function can now be written as  $Q_r={\rm tr}\left[e^{-\beta\hat{H}_0}\right]$ and ${\rm tr}\Bigl[\cdots\Bigr]$ denotes a trace over the nuclear degrees of freedom. 

Wolynes\cite{Wolynes87} made a steepest descent approximation to the time integral in Eq.~(20) to obtain
\begin{equation}
k_{\rm GR} \simeq \frac{\Delta^2}{Q_r\hbar } \sqrt{\frac{2\pi}{-\beta F''(\lambda_{\rm sp})}} e^{-\beta F(\lambda_{\rm sp})}, \label{Wolynes_Theory}
\end{equation}
where $F(\lambda)$ is an effective free energy defined by the equation
\begin{equation}
e^{-\beta F(\lambda)} = \tr[e^{-(\beta-\lambda)\hat{H}_0}e^{-\lambda \hat{H}_1}],
\end{equation}
and $\lambda_{\rm sp}$ is defined by the saddle point condition $F'(\lambda_{\rm sp})=0$. The advantage of the steepest descent approximation is that the resulting rate constant can be evaluated in a straightforward way using imaginary time path integral techniques. Performing a standard $n$-bead path integral discretisation of Eq.~(22) gives\cite{Lawrence18}
\begin{equation}
e^{-\beta F(\lambda_l)} \simeq  \frac{1}{(2\pi\hbar)^{nf}} \int \mathrm{d}^{nf} \mathbf{p} \int \mathrm{d}^{nf} \mathbf{q}\,\,e^{-\beta_n H^{(l)}_{n}(\mathbf{p},\mathbf{q}) },
\end{equation}
where $\beta_n=\beta/n$, $\lambda_l=l\beta_n$, and
\begin{equation}
H^{(l)}_{n}(\mathbf{p},\mathbf{q})\! = h_n(\mathbf{p},\mathbf{q})\! +\!  \sum_{j=0}^{l}w^{(l)}_j V_{1}(\bm{q}_j) + \sum_{j=l}^{n} w^{(l)}_j V_{0}(\bm{q}_j)
\end{equation}
with
\begin{equation}
w^{(l)}_j = 
     \begin{cases}
     0 &\quad\text{if }j=l\,\text{  and  }\,l\in\{0,n\} \\
      \frac{1}{2} &\quad\text{if }j\in\{0,l,n\} \text{ and } l\notin\{0,n\}\\
       1 &\quad\text{otherwise} 
     \end{cases} \label{trapezium_weights}
\end{equation}
and
\begin{equation}
\begin{aligned}
h_n(\mathbf{p},\mathbf{q})= \sum_{j=1}^{n}\sum_{\nu=1}^f \bigg[ \frac{p_{j,\nu}^2}{2 m_{\nu}} + \frac{1}{2} m_\nu \omega_n^2\big(q_{j+1,\nu}-q_{j,\nu}\big)^2\bigg],
\end{aligned}
\end{equation}
where $\omega_n=1/\beta_n\hbar$ and $q_{n,\nu}\equiv q_{0,\nu}$. It follows that the derivatives of the free energy that are needed to evaluate Eq.~(21) can be written as  
\begin{equation}
-\beta F'(\lambda_l) = \Big\langle s(\bm{q}_0)\Big\rangle_{\lambda_l} \label{RP_Fprime},
\end{equation}
and
\begin{equation}
\begin{aligned}
-\beta F''(\lambda_l) = &\Big\langle s(\bm{q}_0)s(\bm{q}_l)\Big\rangle_{\lambda_l}-\Big\langle s(\bm{q}_0)\Big\rangle_{\lambda_l}^2\end{aligned}, \label{RP_F2prime}
\end{equation}
where $s(\bm{q})=V_0(\bm{q})-V_1(\bm{q})$ and
\begin{equation}
\left<A(\mathbf{q})\right>_{\lambda_l} = \frac{\int \mathrm{d}^{nf} \mathbf{p} \int \mathrm{d}^{nf} \mathbf{q}\,\, e^{-\beta_n H^{(l)}_{n}(\mathbf{p},\mathbf{q}) }A(\mathbf{q})}{\int \mathrm{d}^{nf} \mathbf{p} \int \mathrm{d}^{nf} \mathbf{q}\,\, e^{-\beta_n H^{(l)}_{n}(\mathbf{p},\mathbf{q}) }}.
\end{equation}

The simplicity with which these quantities can be evaluated makes it straightforward to apply Wolynes theory to arbitrarily complex (anharmonic and multi-dimensional) condensed phase problems.\cite{Wolynes89,Bader90,Wolynes91} The theory exactly recovers Marcus theory in the high temperature limit for the spin-boson model, and it is known to be very accurate even at extremely low temperatures, where nuclear quantum effects can increase the rate over the Marcus theory prediction by several orders of magnitude.\cite{Lawrence18} It is conceivable that one could do better than the steepest descent approximation and obtain a more accurate imaginary time path integral approximation to the Fermi Golden Rule rate in Eq.~(20), for example along the lines suggested recently by Richardson and co-workers.\cite{Thapa19} However, this would not change the following argument because one could then simply use this improved approximation instead of Wolynes theory to provide the Golden Rule component of the interpolation formula we shall discuss below.

\subsection{RPMD rate theory}

The Born-Oppenheimer approximation to the electron transfer rate can be calculated whenever the reactants (state $\ket{0}$) and products (state $\ket{1}$) can be separated by a position space dividing surface. This is certainly the case for an activated electron transfer reaction in the normal Marcus regime, where the projection operators $\hat{P}_r$ and $\hat{P}_p$ onto the reactant and product states can equally well be replaced by
\begin{equation}
\hat{P}_r = \theta[-s(\hat{\bm{q}})]\phantom{xx} \hbox{and}\phantom{xx} \hat{P}_p = \theta[s(\hat{\bm{q}})],
\end{equation}
where $\theta(x)$ is the Heaviside step function and $s({\bm{q}})=V_0({\bm{q}})-V_1({\bm{q}})$ as above. With such a definition for the dividing surface [at $s({\bm{q}})=0]$, the Born-Oppenheimer rate $k_{\rm BO}$ can be calculated simply by considering the motion of the nuclei on the ground adiabatic potential energy surface
\begin{equation}
U(\bm{q})=\frac{V_0(\bm{q})+V_1(\bm{q})}{2} - \frac{1}{2}\sqrt{(V_0(\bm{q})-V_1(\bm{q}))^2+4\Delta^2}.
\end{equation}

RPMD rate theory\cite{Craig05a,Craig05b} provides an appropriate way to calculate the rate constant on this surface with the inclusion of nuclear quantum effects. In view of Eqs.~(10) to (12), the $n$-bead RPMD approximation to the Born-Oppenheimer rate constant in this context is\cite{Craig07}
\begin{equation}
k_{\rm BO} \simeq \lim_{t\to\infty} {Q_r^{-1}c_{fs}(t)\over 1-[Q_r^{-1}+Q_p^{-1}]\int_0^t c_{fs}(t)\,{\rm d}t},
\end{equation}
where $c_{fs}(t)$ is a RPMD flux-side correlation function
\begin{equation}
\hspace{-0.00125cm}c_{fs}(t) \!\! =\!\! {1\over (2\pi\hbar)^{nf}}\!\!\!\int\!\! {\rm d}^{nf}{\bf p}\!\!\int\!\! {\rm d}^{nf}{\bf q}e^{-\beta_nH_n({\bf p},{\bf q})}\!\dot{\theta}\![s({\bf q})]\!\theta\![s({\bf q}(t))],\!\!\!\!\!\!\!
\end{equation}
$Q_r$ and $Q_p$ are the reactant and product partition functions
\begin{equation}
Q_r = {1\over (2\pi\hbar)^{nf}}\int {\rm d}^{nf}{\bf p}\int {\rm d}^{nf}{\bf q}\,e^{-\beta_nH_n({\bf p},{\bf q})}\theta[-s({\bf q})],
\end{equation}
\begin{equation}
Q_p = {1\over (2\pi\hbar)^{nf}}\int {\rm d}^{nf}{\bf p}\int {\rm d}^{nf}{\bf q}\,e^{-\beta_nH_n({\bf p},{\bf q})}\theta[s({\bf q})],
\end{equation}
$H_n({\bf p},{\bf q})$ is the adiabatic ring polymer Hamiltonian
\begin{equation}
\begin{aligned}
H_n(\mathbf{p},\mathbf{q}) = h_n(\mathbf{p},\mathbf{q}) +  \sum_{j=1}^{n} U(\bm{q}_j),
\end{aligned}
\end{equation}
and $s(\mathbf{q})=0$ is an appropriate dividing surface in the ring-polymer configuration space, the flux through which at time $t=0$ is given by 
\begin{equation}
\dot{\theta}[s(\mathbf{q})] = \delta[s(\mathbf{q})] \sum_{j=1}^{n} \sum_{\nu=1}^{f} \frac{\partial s(\mathbf{q})}{\partial q_{j,\nu} } \frac{p_{j,\nu}}{m_{\nu}}.
\end{equation}

Since the RPMD rate is just a classical rate in the extended phase space of the ring-polymer, it can be calculated using any of the techniques that have been developed over the years for overcoming the rare event problem in the calculation of classical reaction rates. In the test calculations reported below, we have used the Bennett-Chandler method,\cite{Bennett77,Chandler78} which provides a straightforward way to calculate the ratios of $c_{fs}(t)$, $Q_r$, and $Q_p$ that are required in Eq.~(32). RPMD rate theory also has a number of other desirable features, including the following:
\begin{enumerate}
\item It is a full-dimensional theory in which all degrees of freedom are treated on an equal footing. 
\item It is parameter-free: a given reaction at a given temperature has a unique RPMD rate, which can be converged simply by increasing the number of ring-polymer beads.
\item The RPMD rate constant is independent of the choice of the dividing surface,\cite{Craig05b} and it is rigorously consistent with the quantum mechanical equilibrium constant (i.e., the forward and reverse RPMD rates exactly satisfy the quantum mechanical detailed balance condition).
\item The theory becomes exact in the high-temperature (classical) limit, and it is also exact for the shallow tunnelling through a parabolic barrier.\cite{Craig05a}
\item In the low temperature (deep tunnelling) regime, where the classical rate is too small by several orders of magnitude, the RPMD rate is typically within a factor of two of the exact quantum mechanical result. Moreover it provides an approximate  bound on the exact result, because RPMD is known to underestimate the rates of symmetric reactions and to overestimate those of asymmetric reactions. This has been found to be the case for a variety of reactions for which the exact quantum mechanical rates could be computed for comparison,\cite{Collepardo09,Suleimanov11,Perez12,Suleimanov13} and also established theoretically from the connection between RPMD rate theory and semiclassical instanton theory.\cite{Richardson09}
\end{enumerate}
We are not aware of any other electronically adiabatic reaction rate theory that shares all of these desirable features and is also routinely applicable to the calculation of chemical reaction rates in complex (anharmonic and multidimensional) systems such as liquids. 

\section{INTERPOLATION FORMULA} \label{interpol_sec}

In the previous section, we have given expressions for the exact electron transfer rate constant $k(\Delta)$, the Fermi Golden Rule rate constant $k_{\rm GR}(\Delta)$, and the Born-Oppenheimer rate constant $k_{\rm BO}(\Delta)$, all of which can clearly be calculated as a function of the electronic coupling strength $\Delta$. For an electron transfer reaction in solution, Wolynes theory can be used to provide an estimate of $k_{\rm GR}(\Delta)$, and RPMD rate theory an estimate of $k_{\rm BO}(\Delta)$, for any value of $\Delta$. However, there is as yet no equally simple and reliable way to estimate the exact electron transfer rate constant $k(\Delta)$ for values of $\Delta$ away from the non-adiabatic (Golden Rule, $\beta\Delta\ll 1$) and adiabatic (Born-Oppenheimer, $\beta\Delta\gg1$) limits. 

To solve this problem, let us now consider combining the results of the Golden Rule and Born Oppenheimer calculations with the simple interpolation formula
\begin{equation}
        k(\Delta)\simeq k_{\rm IF}(\Delta) = \frac{k_{\rm GR}(\Delta)k_{\rm BO}(\Delta)}{k_{\rm GR}(\Delta)+k_{\rm BO}(\Delta=0)}.\label{interpolation_formula}
\end{equation}
This has the correct limiting behaviour for both small and large $\Delta$, as illustrated in Fig.~1. As $\beta\Delta\to 0$, $k_{\rm BO}(\Delta)$ and $k_{\rm GR}(\Delta)+k_{\rm BO}(\Delta=0)$ both tend to $k_{\rm BO}(\Delta=0)$, which then cancels top and bottom in Eq.~(38) to leave $k_{\rm IF}(\Delta)\simeq k_{\rm GR}(\Delta)$, the correct non-adiabatic rate constant. And when $\beta\Delta\gg1$, $k_{\rm GR}(\Delta)+k_{\rm BO}(\Delta=0)$ tends to $k_{\rm GR}(\Delta)$, which then cancels with the $k_{\rm GR}(\Delta)$ in the numerator to leave $k_{\rm IF}(\Delta)\simeq k_{\rm BO}(\Delta)$, the correct adiabatic rate constant. In effect, the right-hand side of Eq.~(38) is a Pad\'e-like approximant to $k(\Delta)$ which is straightforward to compute even for a liquid phase electron transfer reaction, and which interpolates correctly between the non-adiabatic and adiabatic limits. 

 \begin{figure}[t]
 % \captionsetup{singlelinecheck=off, justification=justified}
 \includegraphics[width=0.4\textwidth]{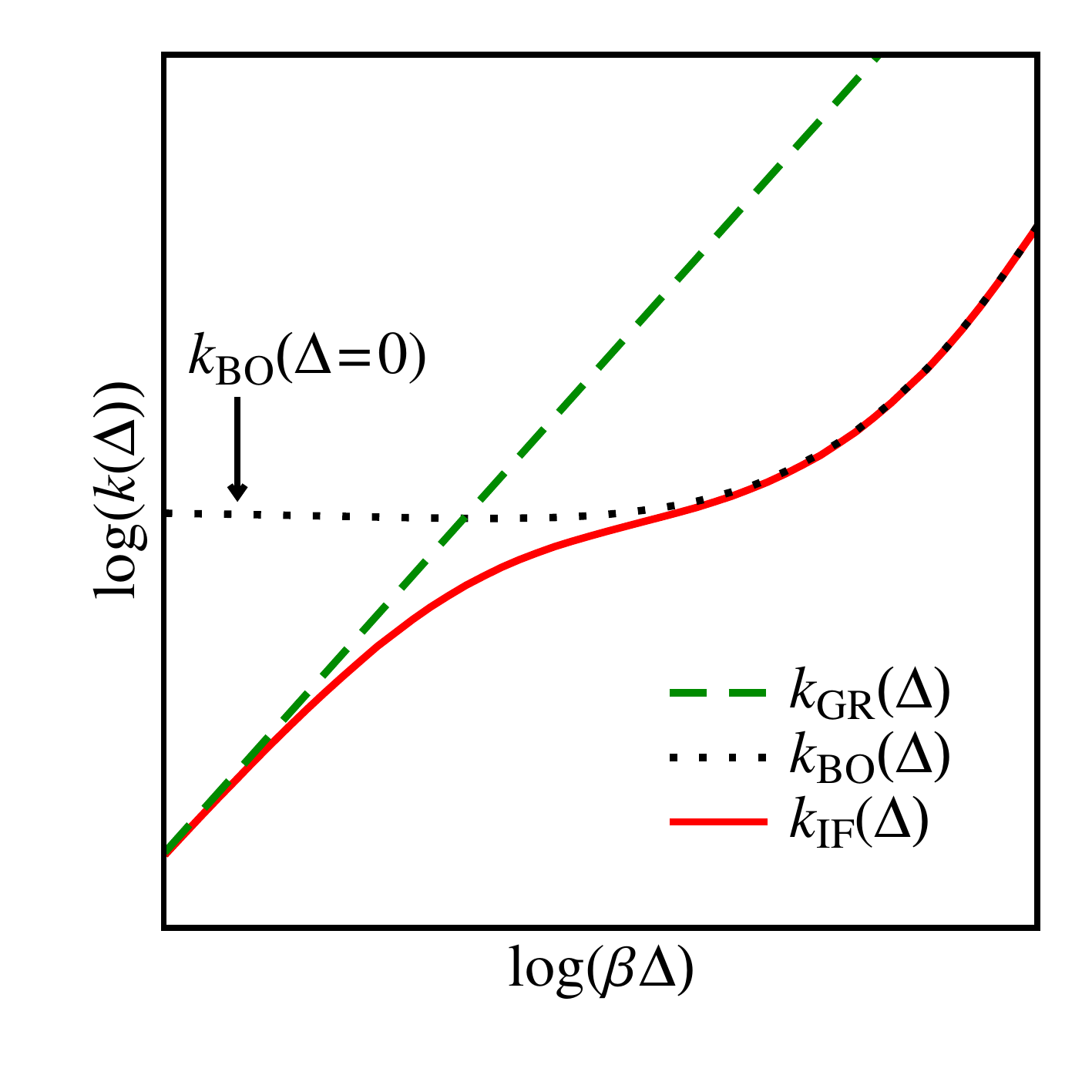}
 \caption{\small Illustration of the behaviour of the interpolation formula in Eq.~(38). For small $\beta\Delta$, $k_{\rm IF}(\Delta)$ approaches the Golden Rule rate constant $k_{\rm GR}(\Delta)$, and for large $\beta\Delta$ it approaches the Born-Oppenheimer rate constant $k_{\rm BO}(\Delta)$, as discussed in more detail the text.}
 \label{Interpolation-Formula}
 \end{figure}

It is clear that the interpolation formula in Eq.~(38) can only be used when the Born-Oppenheimer rate constant is well defined for all relevant values of $\Delta$. So the formula cannot be used in the Marcus inverted regime. And in the normal regime, we require that the reaction barrier is large compared to $k_{\rm B}T$, at least in the non-adiabatic limit. More formally, we require that defining the reactants and products using a position space dividing surface on the lower adiabatic surface is not significantly different to using the diabatic projection operators in the non-adiabatic limit, so that to a good approximation
\begin{equation}
Q_r(\Delta=0) = \tr[e^{-\beta \hat{H}_0}]\simeq  \tr[e^{-\beta \hat{H}_{\rm BO}(\Delta=0)}{\theta}(-s(\hat{\bm{q}}))],
\end{equation}
\begin{equation}
Q_p(\Delta=0) = \tr[e^{-\beta \hat{H}_1}]\simeq \tr[e^{-\beta \hat{H}_{\rm BO}(\Delta=0)}{\theta}(s(\hat{\bm{q}}))], 
\end{equation}   
where $\hat{H}_{\rm BO}(\Delta=0)$ is the Born-Oppenheimer Hamiltonian on the (cusped) lower adiabatic surface with zero electronic coupling. With this expected range of validity in mind we can now consider two important properties of the interpolation formula. 

The first of these properties is that the formula satisfies detailed balance. That is
\begin{equation}
\frac{k_{\rm IF}(\Delta)}{k_{\rm IF}'(\Delta)}=\frac{Q_p(\Delta)}{Q_r(\Delta)},
 \end{equation}
 where $k_{\rm IF}(\Delta)$ and $k_{\rm IF}'(\Delta)$ are the forward and backward rate constants and  $Q_r(\Delta)$ and $Q_p(\Delta)$ are the reactant and product partition functions. This can be seen by first noting that Eq.~(38) gives
 \begin{equation}
\frac{k_{\rm IF}(\Delta)}{k_{\rm IF}'(\Delta)} = {k_{\rm BO}(\Delta)\over k'_{\rm BO}(\Delta)}\cdot{{1+k'_{\rm BO}(\Delta=0)/k'_{\rm GR}(\Delta)}\over{1+k_{\rm BO}(\Delta=0)/k_{\rm GR}(\Delta)}}.
\end{equation}
Provided Eqs.~(39) and~(40) are satisfied, we also have
 \begin{equation}
\frac{Q_p(\Delta=0)}{Q_r(\Delta=0)}={k_{\rm BO}(\Delta=0)\over k'_{\rm BO}(\Delta=0)} =
{k_{\rm GR}(\Delta)\over k'_{\rm GR}(\Delta)},
\end{equation}
because the Born-Oppenheimer and Fermi Golden Rule rates each satisfy detailed balance. Hence
\begin{equation}
{k_{\rm BO}(\Delta=0)\over k_{\rm GR}(\Delta)} = {k'_{\rm BO}(\Delta=0)\over k'_{\rm GR}(\Delta)},
\end{equation}
and combining this with Eq.~(42) gives
 \begin{equation}
 \frac{k_{\rm IF}(\Delta)}{k_{\rm IF}'(\Delta)}= {k_{\rm BO}(\Delta)\over k'_{\rm BO}(\Delta)} = {Q_p(\Delta)\over Q_r(\Delta)}.
\end{equation}

The second important property of Eq.~(38) is that it reduces to a known analytical result, the Zusman equation,\cite{Zusman80} under the appropriate conditions. The Zusman equation gives an analytical expression for the rate constant in the classical limit for the spin-boson model, extending Marcus theory to include the effect of friction along the reaction coordinate. The Zusman equation is given explicitly in Section~\ref{spin_b_sec} but for now we simply note that it can be written in a form reminiscent of Eq.~(\ref{interpolation_formula}) as
\begin{equation}
k_{\mathrm{ZUS}}(\Delta)=\frac{k_{\mathrm{MT}}(\Delta)k_{\mathrm{A}}(\Delta=0)}{ k_{\mathrm{MT}}(\Delta)+k_{\mathrm{A}}(\Delta=0)}, \label{Zusman1}
\end{equation}
where $k_{\mathrm{MT}}(\Delta)$ is the Marcus theory rate and $k_{\mathrm{A}}(\Delta=0)$ is the classical rate on the cusped ground adiabatic potential in the limit of high friction. The obvious difference between this and Eq.~(38), which for the classical spin-boson model becomes
\begin{equation}
k_{\rm IF}(\Delta) = \frac{k_{\mathrm{MT}}(\Delta)k_{\mathrm{A}}(\Delta)}{ k_{\mathrm{MT}}(\Delta)+k_{\mathrm{A}}(\Delta=0)} ,
\end{equation} 
is that it predicts $k_{\rm ZUS}(\Delta)\to k_{A}(\Delta=0)$ rather than $k_{A}(\Delta)$ for large $\Delta$. Hence the Zusman equation misses the fact that increasing $\Delta$ lowers the adiabatic reaction barrier. This is because the derivation of the Zusman equation assumes that $\beta\Delta\ll1$. Under these circumstances $k_A(\Delta)\simeq k_A(\Delta=0)$, and the classical limit of our interpolation formula reduces to the Zusman equation.

 \color{black} The breakdown of the Zusman equation for large electronic coupling is well known, and there already exist classical theories which can be used to calculate rates in this regime.\cite{Calef83,Pollak90,Rips95,Starobinets96,Gladkikh05} Our approach is closely related to several of these theories when it is applied to the high temperature limit of the spin-boson model. In particular, we note that Eq.~(47) is very similar to an interpolation formula proposed by Gladkikh \emph{et al.},\cite{Gladkikh05} which corrects the Zusman equation by interpolating between the rates for the cusped and parabolic barriers in the adiabatic limit. The present approach can thus be viewed as a generalisation of these previous interpolation formulas which is capable of treating general anharmonic condensed phase problems and including the effects of tunnelling and zero point energy.    \color{black}
 
 \begin{figure}[!t]
%  \captionsetup{singlelinecheck=off, justification=raggedright}
 \resizebox{1.0\columnwidth}{!} {\includegraphics{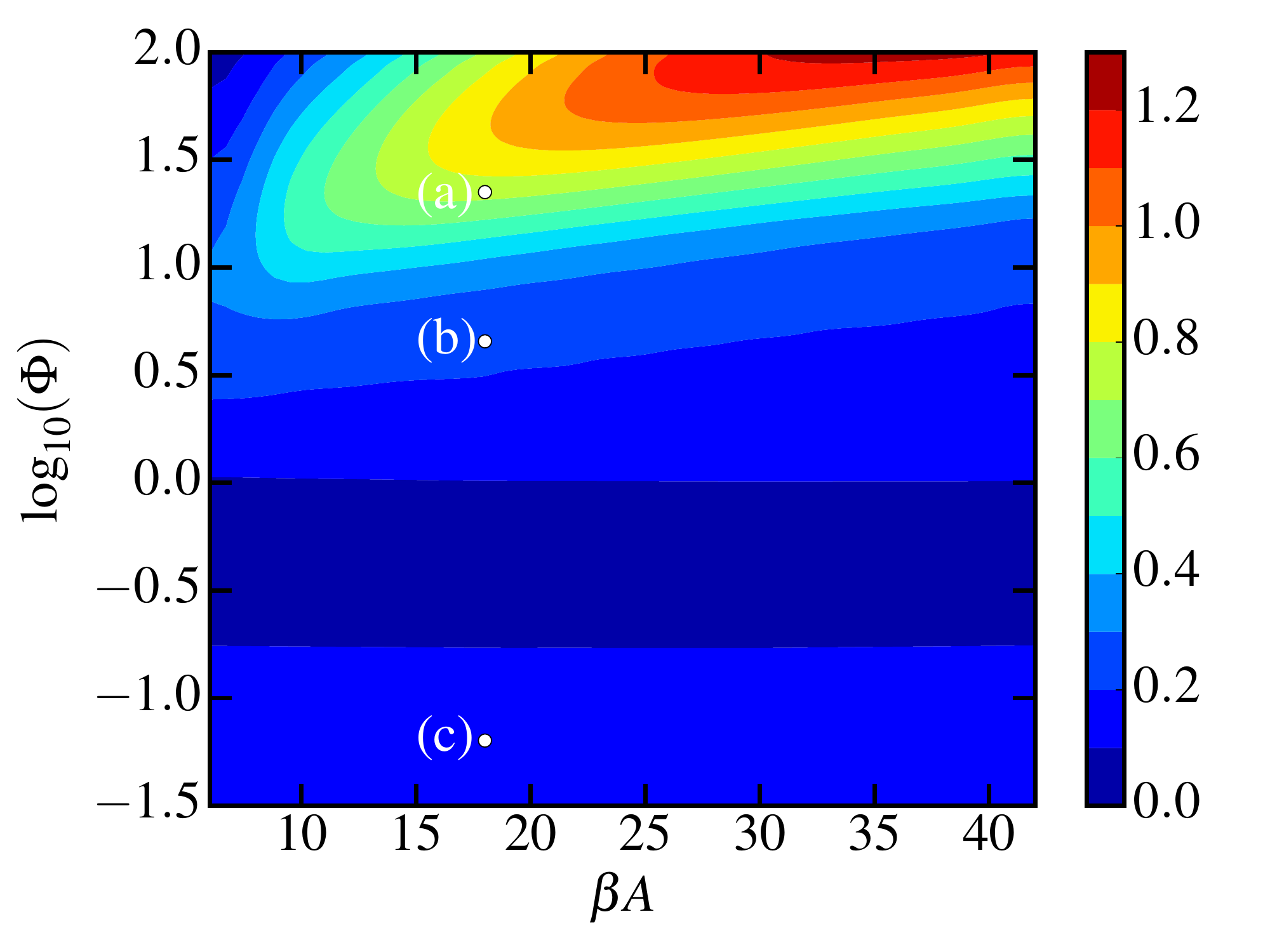}}
 \centering
 \caption{\small Contour plot of the maximum error as defined in Eq.~(\ref{Error_Def}) for the scattering model. The exact rate $k(\Delta)$ was calculated using the log derivative method, as were $k_{\rm GR}(\Delta)$ and $k_{\rm BO}(\Delta)$. } 
 \label{contour_plot1}
 \end{figure}
 
  \begin{figure*}[t]
% \captionsetup{singlelinecheck=off, justification=raggedright}
 \resizebox{0.9\textwidth}{!} {\includegraphics{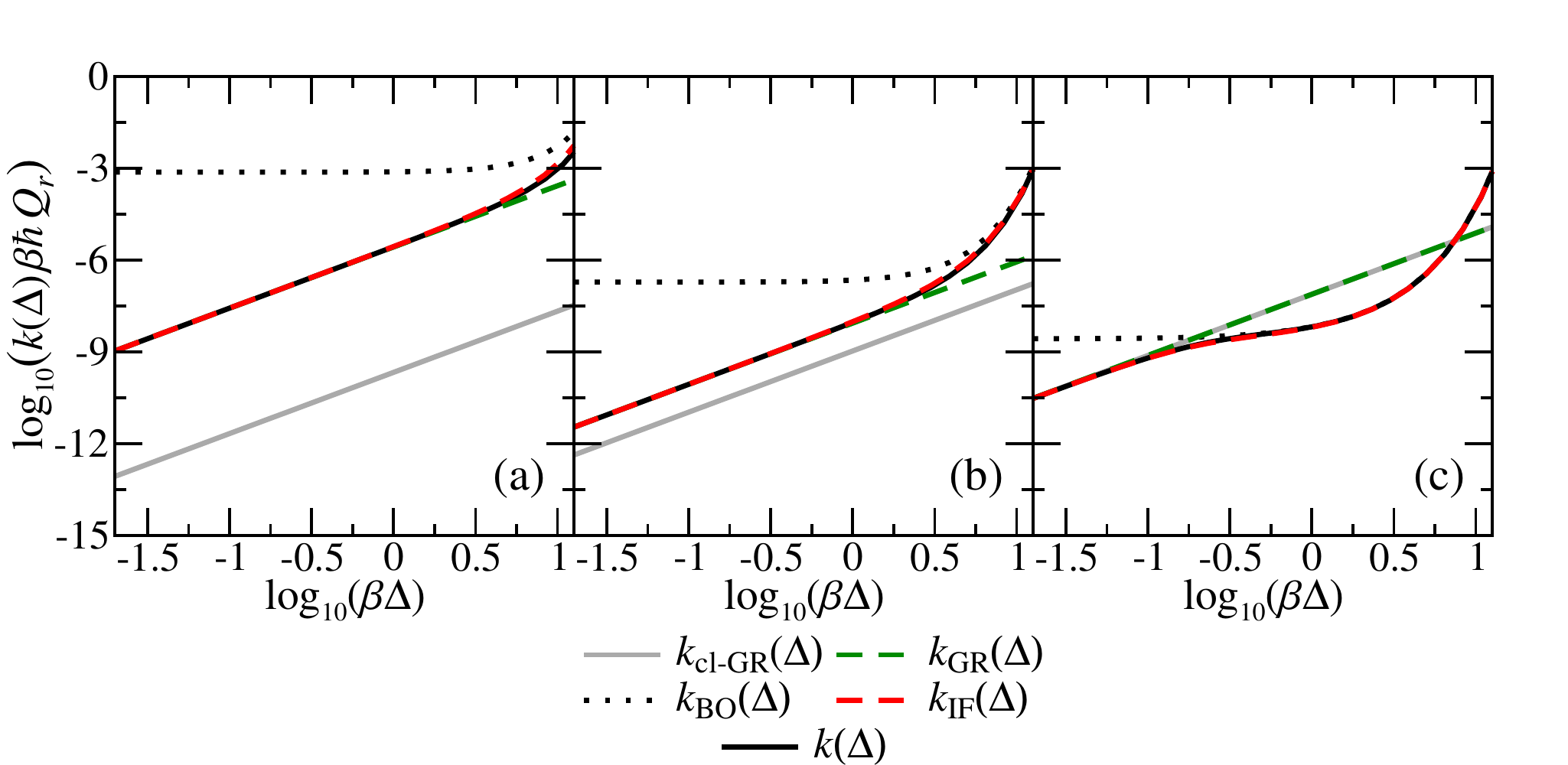}}
 \caption{\small Plots demonstrating the accuracy of the interpolation formula for the three points labeled in Fig.~\ref{contour_plot1}. All three systems have $\beta A=18$, with (a) $\Phi=22.4$, (b) $\Phi=4.54$, and (c) $\Phi=0.0635$. The exact rate $k(\Delta)$ was calculated using the log derivative method, as were $k_{\mathrm{GR}}(\Delta)$ and $k_{\mathrm{BO}}(\Delta)$. The classical Golden Rule rate was calculated using Eq.~(\ref{LZRATE}).}
 \label{contour_examples}
 \end{figure*}
 
\section{Scattering Model} \label{Scattering_Section}

As a first test of the accuracy of the interpolation formula in Eq.~(38), we shall consider a simple one-dimensional curve crossing model with the following diabatic potential energy curves:
\begin{equation}
V_0(q) = A e^{+q/L},
\end{equation}
\begin{equation}
V_1(q) = A e^{-q/L} .
\end{equation}
This provides a useful test problem because not only can we calculate the exact rate to go from state $\ket{0}$ to $\ket{1}$, but we can also calculate each of the components of the interpolation formula exactly. The exact rate $k(\Delta)$ and the ground adiabatic Born-Oppenheimer rate $k_{\rm BO}(\Delta)$ can each be calculated as Boltzmann averages of appropriate cumulative reaction probabilities,
\begin{equation}
k(\Delta) = {1\over (2\pi\hbar)Q_r}\int_0^{\infty} e^{-\beta E}N(E)\,{\rm d}E,
\end{equation}
where $Q_r=\sqrt{m/2\pi\beta\hbar^2}$ is the reactant partition function per unit length. The cumulative reaction probability $N(E)$ that determines the exact non-adiabatic rate $k(\Delta)$ can be calculated using the coupled channel log derivative method,\cite{Johnson73} and the Golden Rule rate $k_{\rm GR}(\Delta)$ can be extracted from the $\Delta^2$ dependence of $k(\Delta)$ in the limit as $\Delta\to 0$. For the Born-Oppenheimer rate $k_{\rm BO}(\Delta)$, $N(E)$ can be calculated using a single channel log derivative method on the ground adiabatic potential energy surface.  The fact that all three rates $k(\Delta)$, $k_{\rm GR}(\Delta)$ and $k_{\rm BO}(\Delta)$ can be calculated exactly allows us to test the accuracy of the interpolation formula in Eq.~(38) separately from that of the approximate path integral techniques (Wolynes theory and RPMD rate theory) which are required for condensed phase problems. 

 The behaviour of this model system is completely determined by three dimensionless parameters: $\beta \Delta$, $\beta A$, and ${mL^2}/{\beta \hbar^2}$. Keeping the last two of these parameters fixed we define the error in the interpolation formula as a function of $\beta\Delta$ using the equation
\begin{equation}
E(\beta \Delta) =  \bigg|\log_2\bigg(\frac{k_{\rm IF}(\beta\Delta)}{k(\beta\Delta)}\bigg)\bigg|.
\end{equation}
This allows us to define the maximum error for a particular $\beta A$ and ${mL^2}/{\beta \hbar^2}$ as
\begin{equation}
E_{\max} = \max_{\beta\Delta<0.8 \beta A} E(\beta\Delta),
\label{Error_Def}
\end{equation}
where the range of $\beta \Delta$ is limited to ensure that the reaction is always activated. 
This is a more useful measure of the error than the percentage error as it treats underestimation and over estimation on an equal footing, and is in keeping with the usual way that errors are discussed for RPMD, which is expected to be within a factor of 2 of the exact rate in the deep tunnelling regime. Note that $E_{\max}=1$ corresponds to the interpolation formula being out by at most a factor of 2 and $E_{\max}=0$ corresponds to it being exact for all values of $\beta \Delta$.

Figure \ref{contour_plot1} shows $E_{\max}$ as a function of the two remaining free parameters,  which we take to be $\beta A$ and the dimensionless parameter $\Phi$ defined as
\begin{equation}
\Phi =  \beta A \sqrt{2\beta\hbar^2\over \pi mL^2}.
\end{equation}
This parameter can be related to the instanton action for a linear approximation to the diabatic potentials in the non-adiabatic limit and it quantifies how \lq\lq quantum mechanical" the reaction is. We see that in the classical regime,  $\log_{10}(\Phi)<0$, the error is essentially solely a function of $\Phi$ and approaches a plateau with $E_{\max}$ just above 0.1, corresponding to the interpolation formula being within a factor of $1.1$ of the exact rate for the full range of $\beta \Delta$. For $\log_{10}(\Phi)>0$, the error grows as the problem becomes more quantum mechanical. However $E_{\max}$ remains below 1 for $\log_{10}(\Phi)<1.5$, corresponding to the interpolation formula being within a factor of 2 of the exact rate throughout this regime. As an illustration of the physical regimes that the plot spans, note that the point labelled (a) corresponds to a strongly quantum mechanical system with $\beta A=18$, $T=50\, \mathrm{K}$, $L=0.5\,{a}_0$, and $m=1000\,m_{\rm e}$, whereas the point labelled (c) corresponds to a system in which the mass associated with the nuclear motion has been increased to $m=19000\,m_{\rm e}$.

Figure \ref{contour_examples} shows the rate as a function of $\beta \Delta$ for the three labelled points in figure \ref{contour_plot1}. The plot shows the exact rate as well as the result of the interpolation formula evaluated using the exact Born-Oppenheimer and Fermi Golden Rule rates. To illustrate the importance of nuclear quantum effects we also include the classical limit of the Golden Rule rate which is given by\cite{Nitzan06}
\begin{equation}
k_{\text{cl-GR}} = \frac{\Delta^2 }{\hbar^2 Q_r} \sqrt{\frac{\pi m}{2\beta}}\frac{ L}{ A} e^{-\beta A}. \label{LZRATE}
\end{equation}
We see that the interpolation formula is capable of capturing the full range of behaviour, from the highly quantum mechanical in system (a) to the essentially classical in system (c). It is also of note that as $\Phi$ decreases the value of $\beta\Delta$ at which the reaction begins to behave adiabatically also decreases. This can be understood by noting that $\Phi$ is related to the Landau-Zener adiabaticity parameter\cite{Nitzan06}
\begin{equation}
P(\Delta) = \frac{2\pi\Delta^2}{\hbar \langle|v|\rangle_{\mathrm{cl}}|\delta F|}
\end{equation}
where $\langle|v|\rangle_{\mathrm{cl}}$ is the classical average thermal velocity,
\begin{equation}
\frac{1}{2}\langle|v|\rangle_{\mathrm{cl}}=\frac{1}{\sqrt{2\pi m\beta}},
\end{equation}
and $\delta F= 2A/L$. We see that $\Phi=1/P(k_{\rm B}T)$ and so gives a measure of how non-adiabatic the reaction is expected to be at $\beta\Delta=1$. In system (a), $\Phi\gg1$, and the rate is firmly non-adiabatic at $\beta\Delta=1$, whereas in system (c), $\Phi\ll 1$, and the rate is firmly adiabatic at $\beta\Delta=1$.  This connection between $\Phi$ (or equivalently the instanton action) and the Landau-Zener adiabaticity parameter can be thus be used to account for the fact that increased nuclear tunnelling tends to make a reaction more non-adiabatic.

\section{Spin-boson model} \label{spin_b_sec}

The spin-boson model is the prototypical non-adiabatic system that is used to model electron transfer reactions in condensed phase environments. Written in the ``reaction coordinate'' form, the diabatic potentials of this model are given by\cite{Garg85}
\begin{equation}
V_0(Q,\bm{q})=\frac{1}{2}\Omega^2 \Bigg(Q + \sqrt{\frac{\Lambda}{2\Omega^2}}\Bigg)^2 +  V_{sb}(Q,\bm{q}),\label{reaction-coorda}
\end{equation}
\begin{equation}
V_1(Q,\bm{q})=\frac{1}{2}\Omega^2 \Bigg(Q - \sqrt{\frac{\Lambda}{2\Omega^2}}\Bigg)^2  +  V_{sb}(Q,\bm{q})-\epsilon,\label{reaction-coordb}
\end{equation}
where
\begin{equation}
V_{sb}(Q,\bm{q})=\sum_{\nu=1}^{N_b} \frac{1}{2}\omega_\nu^2\Bigg(q_\nu-\frac{c_\nu Q}{\omega_\nu^2}\Bigg)^2,\label{reaction-coordc}
\end{equation} 
and $\Lambda$ is the Marcus theory reorganisation energy (note that we shall work throughout this section with mass-weighted coordinates). The influence of the bath on the reaction coordinate, $Q$, is fully described by the associated spectral density, which is formally defined as
\begin{equation}
J(\omega)=\frac{\pi}{2}  \sum_{\nu=1}^{N_b} \frac{c_{\nu}^2}{{\omega}_{\nu}} \delta(\omega-\omega_\nu).
\end{equation}
The bath is then taken to be infinite, corresponding to a continuous spectral density. Here we choose to consider a purely Ohmic spectral density,
\begin{equation}
J(\omega)= \gamma \omega,
\end{equation}
which gives rise to Langevin dynamics along the reaction coordinate in the classical adiabatic limit. This model problem allows us to investigate the accuracy of our newly proposed method in a variety of different regimes, from underdamped ($\gamma<2\Omega$) to overdamped ($\gamma>2\Omega$) motion along the reaction coordinate, from high frequencies (large $\Omega$) to low frequencies (small $\Omega$), and for the full range of non-adiabatic (small $\Delta$) to adiabatic (large $\Delta$) behaviour.

The reaction coordinate form can be related to the conventional spin-boson model by a normal mode transformation.\cite{Garg85,Leggett84,Thoss01} In the conventional form, the diabatic potentials are just sums of uncoupled displaced harmonic oscillators,  
\begin{equation}
V_0(\bm{x})=\sum_{\nu=0}^{N_{b}} \frac{1}{2}\bar{\omega}_\nu^2\Big(x_\nu+\frac{\bar{c}_\nu}{\bar{\omega}_\nu^2}\Big)^2,
\end{equation}
\begin{equation}
V_1(\bm{x})=\sum_{\nu=0}^{N_{b}} \frac{1}{2}\bar{\omega}_\nu^2\Big(x_\nu-\frac{\bar{c}_\nu}{\bar{\omega}_\nu^2}\Big)^2-\epsilon,
\end{equation}
where  $x_\nu$, $\bar{\omega}_\nu$ and $\bar{c}_\nu$  are the transformed (mass weighted) coordinates, frequencies and couplings respectively. In order to map between the two forms one simply needs to know the relation between the spectral density for the coupling of nuclear coordinates to the electronic state
\begin{equation}
J_\sigma(\omega)=\frac{\pi}{2}  \sum_{\nu=0}^{N_b} \frac{\bar{c}_{\nu}^2}{\bar{\omega}_{\nu}} \delta(\omega-\bar{\omega}_\nu),
\end{equation}
and the spectral density $J(\omega)$ for the coupling of the bath to the reaction coordinate. This mapping is given by\cite{Garg85,Leggett84,Thoss01}
\begin{equation}
J_\sigma(\omega)= \frac{\Lambda}{2}\frac{ \Omega^2 J(\omega)}{(\omega^2-\Omega^2+R(\omega))^2+J^2(\omega)},
\end{equation}
where
\begin{equation}
R(\omega)=\frac{2\omega^2}{\pi}P\!\!\int_0^\infty \frac{ J(u)}{u(u^2-\omega^2)} du,
\end{equation}
and $P$ denotes the Cauchy principal value. For the Ohmic spectral density considered here, $R(\omega)=0$, and the electronic spectral density becomes
\begin{equation}
J_\sigma(\omega)= \frac{\Lambda}{2}\frac{\gamma\, \Omega^2 \omega}{(\omega^2-\Omega^2)^2+\gamma^2\omega^2}, \label{SB_spectral_density}
\end{equation}
which is typically referred to as the Brownian oscillator spectral density.

 \begin{figure*}[!t]
    \centering
    \begin{minipage}{.5\textwidth}
        \centering
        \includegraphics[width=1.0\linewidth]{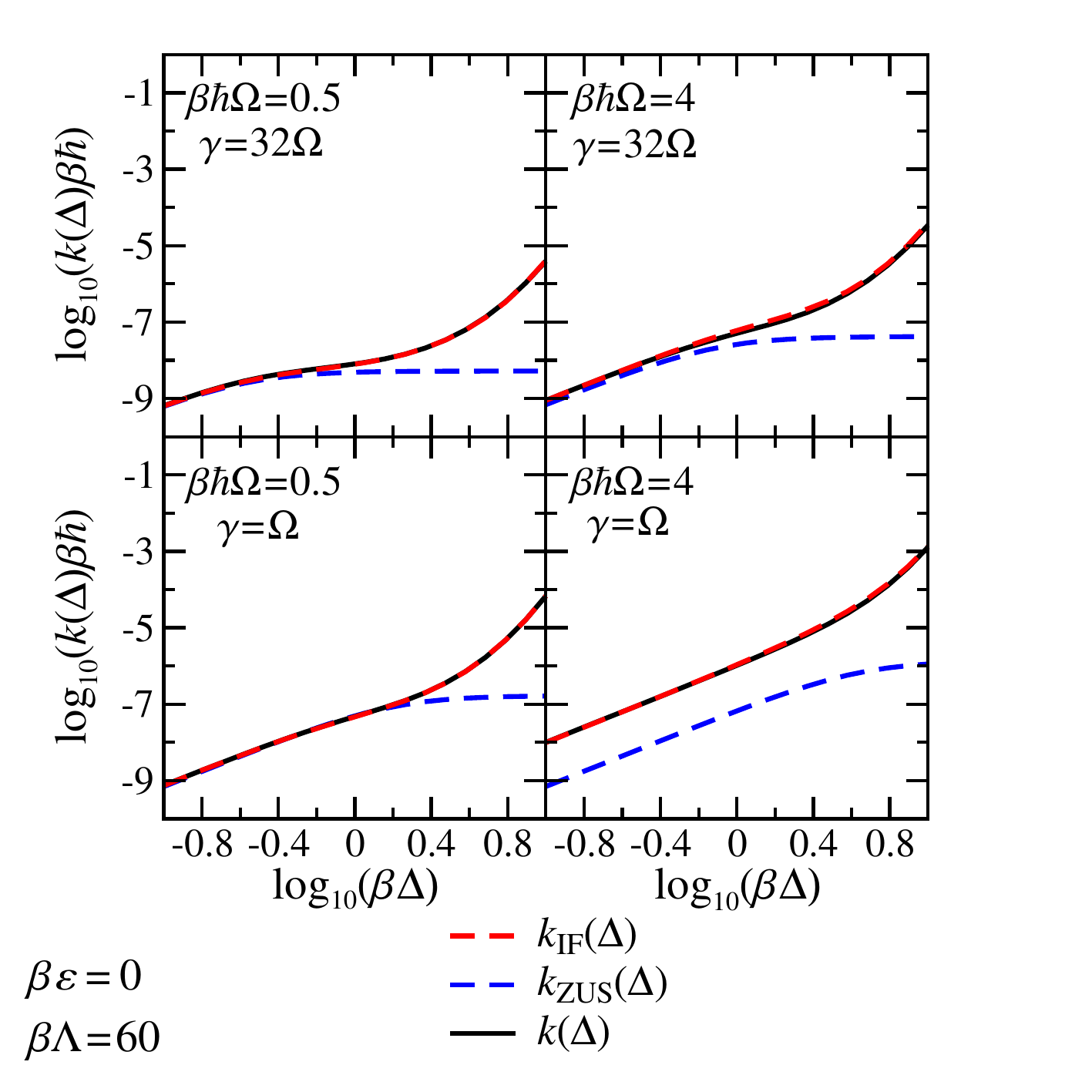}
        (a) Symmetric electron transfer
        \label{fig:prob1_6_2}
    \end{minipage}%
    \begin{minipage}{0.5\textwidth}
        \centering
        \includegraphics[width=1.0\linewidth]{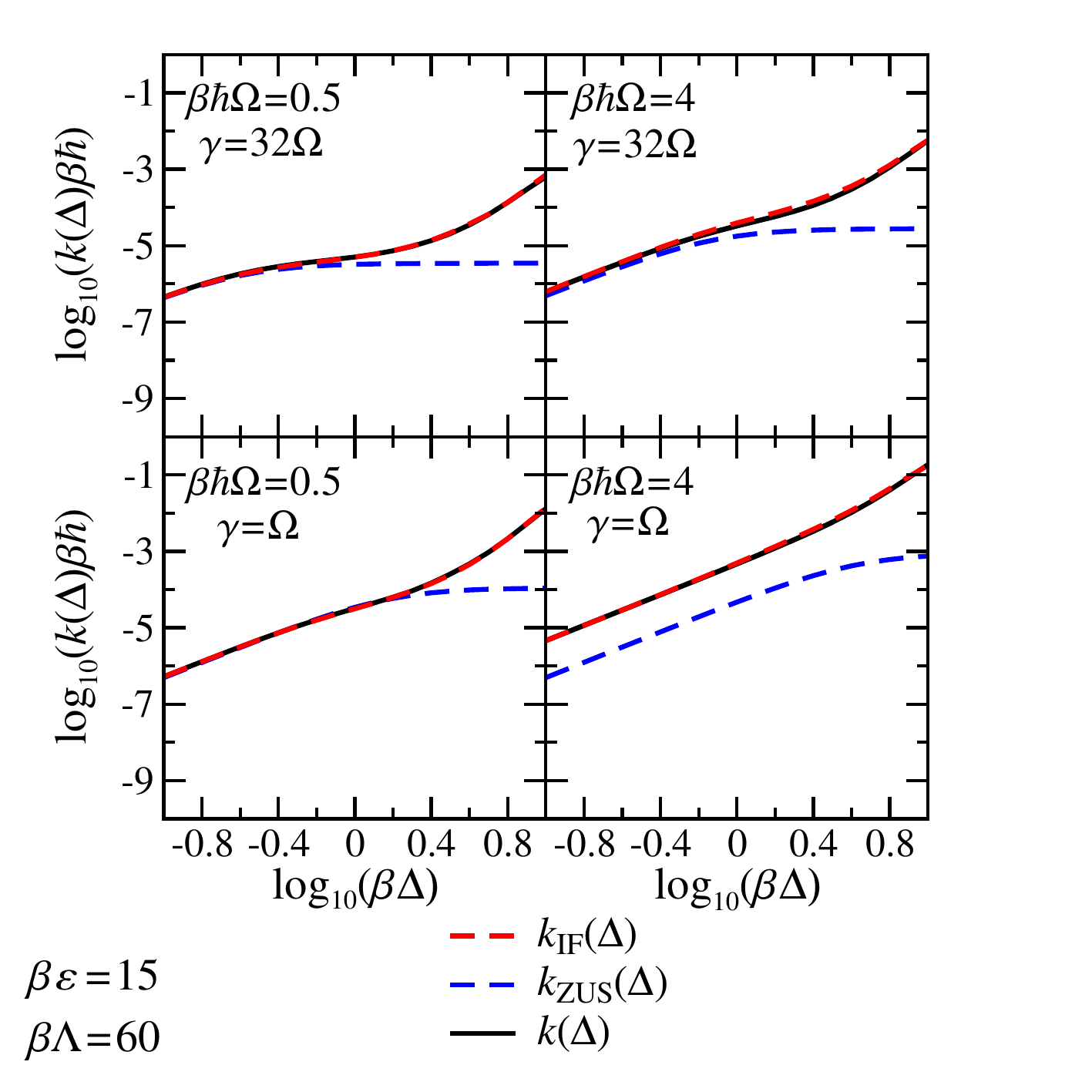}
        (b) Asymmetric electron transfer
        \label{fig:prob1_6_1}
    \end{minipage}
    \caption{\small Plot comparing the rates predicted by the interpolation formula [Eq.~(38)], Zusman theory [Eq.~(70)], and the numerically exact HEOM method, for the spin-boson model with Brownian oscillator spectral density [Eq.~ (\ref{SB_spectral_density})]. The Born-Oppenheimer and Golden Rule rates used in the interpolation formula were calculated using RPMD rate theory and Wolynes theory, respectively.}
    \label{Fig4}
\end{figure*}

The spin-boson model provides a useful system for testing new methods because it is possible to obtain exact numerical results for this model for comparison. Despite this, previous attempts to extend RPMD to treat non-adiabatic rates have only been tested in certain limiting regimes, for example by comparison with Zusman theory or the quantum Golden Rule rate. Here we test the validity of Eq.~(\ref{interpolation_formula}) by comparison with numerically exact results generated using the hierarchical equations of motion (HEOM) method,\cite{Tanimura89,Tanimura90,Tanimura06,Yan04,Xu05,Xu07,Jin08} which allows us to examine a wider range of physically relevant regimes. 

In order to calculate the rate using HEOM we again make use of Eq.~(10) with $\hat P_r=\dyad{0}{0}$ and $\hat P_p = \dyad{1}{1}$. Since it is computationally impractical to prepare the initial density matrix in the form of Eq.~(3), we instead follow the procedure of Shi \emph{et al.},\cite{Shi09} in which the system is initially prepared on the reactant diabat with the bath in thermal equilibrium on the average diabatic potential, and is then equilibrated in the absence of coupling ($\Delta=0$). The resulting initial condition can be expressed as
\begin{equation}
\hat{\rho}(0)=\lim_{t\to\infty}\frac{e^{-i\hat{H}_0t/\hbar}e^{-\beta \hat{\bar H} }\dyad{0}{0}e^{+i\hat{H}_0t/\hbar}}{\Tr[e^{-\beta \hat{\bar H} }\dyad{0}{0}]}.
\end{equation} 
where
\begin{equation}
 \hat{\bar H} = \frac{\hat{H}_0+\hat{H}_1}{2}.
\end{equation}
With this modified initial condition the time-dependent populations of the two electronic states again eventually settle down to conform to Eqs.~(7) and~(8), allowing the exact HEOM forward electron transfer rate to be calculated from Eq.~(10).
 
To highlight the importance of nuclear quantum effects, and the effect of the changes in shape and height of the adiabatic barrier with $\Delta$, we also compare the results of the interpolation formula to the Zusman equation, which for the spin-boson model as described here is\cite{Zusman80,Hartmann00}
\begin{equation}
k_{\mathrm{ZUS}}(\Delta)=\frac{\beta\frac{\Delta^2}{\hbar} \frac{\Omega^2}{4\gamma}\Big(1-\frac{\epsilon^2}{\Lambda^2}\Big)\exp[-\beta\frac{(\Lambda-\epsilon)^2}{4\Lambda}]}{ \frac{\Delta^2}{\hbar}\sqrt{\frac{\pi\beta}{\Lambda}}+\frac{\Omega^2}{4\gamma}\sqrt{\frac{\beta\Lambda}{\pi}}\Big(1-\frac{\epsilon^2}{\Lambda^2}\Big)}. \label{Zusman2}
\end{equation}
This can be seen to be identical to Eq.\ (\ref{Zusman1}) with the identifications 
\begin{equation}
k_{\mathrm{MT}}(\Delta)=\frac{\Delta^2}{\hbar}\sqrt{\frac{\pi\beta}{\Lambda}}\exp[-\beta\frac{(\Lambda-\epsilon)^2}{4\Lambda}],
\end{equation}
and
\begin{equation}
k_{\mathrm{A}}(\Delta=0)=\frac{\Omega^2}{4\gamma}\sqrt{\frac{\beta\Lambda}{\pi}}\Big(1-\frac{\epsilon^2}{\Lambda^2}\Big)\exp[-\beta\frac{(\Lambda-\epsilon)^2}{4\Lambda}].
\end{equation}

In the following we use RPMD to calculate the Born-Oppenheimer rate constant, $k_{\mathrm{BO}}(\Delta)$, and Wolynes theory to calculate the Golden Rule rate constant, $k_{\mathrm{GR}}(\Delta)$, \color{black} although we reiterate that the interpolation formula can be used with any methods which are applicable to these two limits. \color{black} Previous path integral studies of system-bath models have opted to discretise the spectral density using a procedure such as that described in the original RPMD rate theory paper,\cite{Craig05a} increasing the number of modes in the discretisation until the results were converged to the infinite bath limit. Here instead we note that, working in the reaction coordinate form, we can analytically integrate out all but one nuclear degree of freedom.\cite{Cortes85} This results in an effective renormalised ring-polymer Hamiltonian for the single remaining nuclear coordinate (the reaction coordinate, $Q$), which makes the calculations trivial. In the case of RPMD the dynamics of the full system plus bath are then recovered using a Generalised Langevin Equation (GLE). Full details of the derivation of this GLE and the renormalised Hamiltonian, along with how they are implemented numerically, are given in the appendix. \color{black} In principle we could apply the same procedure to calculate the Wolynes theory rates, but for the spin-boson model there is no need since $F(\lambda)$ can be written directly in terms of an integral of $J_\sigma(\omega)$.\cite{Weiss08} \color{black}

Figure~\ref{Fig4} compares the rates obtained using the interpolation formula with the exact HEOM results for the spin-boson model described above. In these calculations we used a typical Marcus reorganisation energy of $\Lambda=60\,k_{\mathrm{B}}T$, with a series of different driving forces ($\epsilon$), solvent frictions ($\gamma$), and reaction coordinate frequencies ($\Omega$). The interpolation formula is seen to give excellent agreement with the exact results in all of the regimes considered, with the largest error, of just under $30\%$, observed in the overdamped high frequency systems \color{black} at intermediate values of $\beta\Delta$ \color{black} ($\beta\hbar\Omega=4$, $\gamma=32\Omega$, $\beta\epsilon=0\text{ and }15$). The interpolation formula is also seen to be most accurate for the overdamped low frequency systems. This is to be expected from its connection with the Zusman equation (see Section~III), which is derived in the limit of low frequency and high friction. It is of note, however, that the interpolation formula is much more accurate than the Zusman equation at even moderate values of $\beta\Delta$, due to the failure of Zusman theory to capture the effect of $\Delta$ on the adiabatic reaction barrier. This breakdown illustrates the importance of testing new methods against numerically exact results rather than relying on comparisons to analytical formulas which can turn out to have limited accuracy. \color{black} (There are of course more accurate formulas than the Zusman formula for the classical limit of the spin-boson model,\cite{Rips95,Gladkikh05} as we have already mentioned in Secs.~I and~III. However, even these are no substitute for the numerically exact HEOM results we have compared to here.) \color{black} 

 The Zusman equation also fails to capture the effects of nuclear tunnelling and zero point energy, which for the underdamped high frequency systems lead to over an order of magnitude increase in the rate constant in the non-adiabatic limit. Since both Wolynes theory and RPMD are able to accurately capture the effect of tunnelling and zero point energy, so too is the interpolation formula when these two theories are used as input. The success of the interpolation formula in this regime (large $\Omega$ and underdamped) is particularly encouraging as it is the opposite of the regime where the Zusman equation is valid. We note that, as was seen in Section~\ref{Scattering_Section}, the greater nuclear tunnelling in this regime means that the Golden Rule expression for the rate continues to be accurate even for $\beta\Delta>1$. The much smaller increase in the rate observed for the overdamped high frequency systems illustrates the reduction of nuclear tunnelling due to friction.\cite{Leggett84} This can be understood in the ring-polymer perspective by considering the renormalisation of the normal mode frequencies of the ring-polymer due to friction, $\omega_k\to\sqrt{\smash{\omega_k^2}+\gamma\omega_k\vphantom{)}}$ (see the appendix). It follows from this renormalisation that a larger value of $\gamma$ leads to a smaller radius of gyration of the ring-polymer and a more classical-like dynamics. 
 
  \begin{figure}[t]
 %\captionsetup{singlelinecheck=off, justification=raggedright}
 \resizebox{0.4\textwidth}{!} {\includegraphics{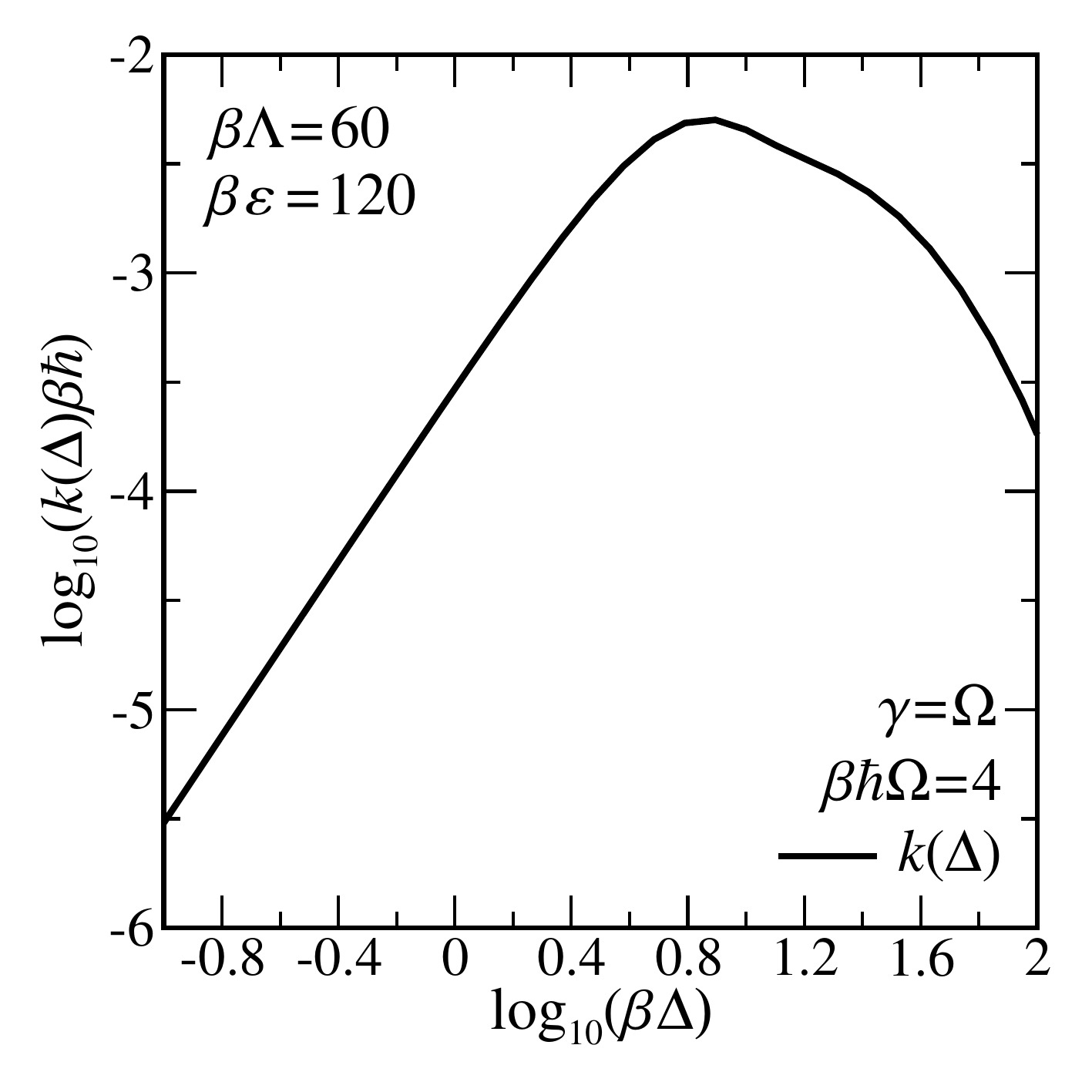}}
 \centering
 \caption{\small Plot of the exact HEOM rate versus $\beta\Delta$ for a spin-boson model with a Brownian oscillator spectral density, deep in the Marcus inverted regime ($\epsilon=2\Lambda$). The interpolation formula in Eq.~(38) cannot be used for this problem because the Born-Oppenheimer rate is not well-defined.}
 \label{Fig6}
 \end{figure}

The success of the interpolation formula in such a wide range of parameter regimes is clearly encouraging. However, it is important to remember its limitations. The formula it is only expected to be accurate when the reactants and products can equally well be defined using either a position space dividing surface or diabatic state projection operators in the non-adiabatic limit. In the most extreme case of the Marcus inverted regime there is no position space description of the reaction and a Born-Oppenheimer rate cannot be defined. Figure~\ref{Fig6} illustrates the behaviour of the exact HEOM rate as a function of $\beta\Delta$ for an underdamped high frequency system with a driving force of $\epsilon=2\Lambda$, deep inside the inverted regime. At small $\beta\Delta$ the rate is well described by the Golden Rule, but at large $\beta\Delta$ the rate begins to decrease as it approaches an adiabatic limit in which there is no transfer of population between the upper and lower adiabats. This kind of system clearly cannot be studied with the interpolation formula in Eq.~(38). We have included it here to provide both an honest illustration of one of the limitations of the current approach and a challenge for future work. 

Despite this limitation we would like to stress that the interpolation formula can still be very accurate even when the barrier to reaction is low. For example, in Fig.~\ref{Fig4}b the barrier on the lower adiabatic surface becomes very small as $\beta\Delta$ becomes large, dropping all the way to $1.2\,k_{\mathrm{B}}T$ when $\beta\Delta=10$. But the rate predicted using the interpolation formula still agrees very well with the exact rate calculated using HEOM with diabatic state projection operators.

\section{Conclusion}
In this paper, we have shown how the RPMD rate theory and Wolynes theory approximations to Born-Oppenheimer and Fermi Golden Rule rates can be combined in a simple interpolation formula that accurately predicts the rates of non-adiabatic reactions with arbitrary electronic coupling strengths. The accuracy of the interpolation formula has been demonstrated by comparison with exact results for both a simple one-dimensional scattering problem and a demanding series of spin-boson models. In particular, the use of the exact HEOM method to provide benchmark results for these spin-boson models has allowed us to explore a wide range of chemically relevant regimes, from underdamped to overdamped dynamics, small to large nuclear quantum effects, and the full range of electronic coupling strengths, from the non-adiabatic to adiabatic limits. Previous methods have not been tested in such wide ranging and challenging regimes. We hope that the exact results presented here will prove a useful test set against which future methods can be evaluated. It would also be interesting to see some of the existing methods which have been proposed for calculating non-adiabatic reaction rates tested against the same set of problems, rather than compared with simple analytical formulas for the rate which have limited ranges of applicability. In order to facilitate this, we have tabulated the numerical data we used to produce Fig.~4 in the supplementary material.

Since RPMD rate theory and Wolynes theory are both readily applicable to truly complex (anharmonic and multidimensional) problems, it will be interesting in future work to apply the interpolation formula to more realistic models of  electronically non-adiabatic reactions and use it to answer some chemically interesting questions. We note in particular that the formula is well suited to quantifying the degree of electronic non-adiabaticity in a reaction, as well as assessing the relative importance of electronic coupling and solvent friction in driving a reaction away from the non-adiabatic limit, as we have shown in some of our examples. Further qualitative insights into the reaction mechanism can also clearly be gained from the Born-Oppenheimer and Golden Rule calculations that the interpolation formula combines.

Having said this, there are clearly still some open issues which warrant further theoretical work. In addition to the inability of the interpolation formula to treat systems in the Marcus inverted regime, we still need to investigate its accuracy for systems in which the Condon approximation breaks down, and for systems such as proton-coupled electron transfer reactions which can have multiple competing reaction pathways. Further work is also needed in order to treat systems in which multiple electronic states are involved, such as in superexchange mediated electron transfer, and to develop approaches that can treat both the incoherent rate processes considered here and also the coherent dynamics relevant to condensed phase electronic spectra, which current non-adiabatic methods such as Wolynes theory cannot describe.

\section*{supplementary material}
In the supplementary material, we tabulate the RPMD rate theory, Wolynes theory, interpolation formula, and numerically exact HEOM rates we used to prepare each of the eight panels in Fig.~4.

\begin{acknowledgments}
We would like to thank Thomas P. Fay for helpful discussions, and Jeremy Richardson for his comments on the first draft of this manuscript. J. E. Lawrence is supported by The Queen's College Cyril and Phillis Long Scholarship in conjunction with the Clarendon Fund of the University of Oxford and by the EPRSC Centre for Doctoral Training in Theory and Modelling in the Chemical Sciences, EPSRC grant no. EP/L015722/1. T. Fletcher is also supported by this EPSRC grant, and L. P. Lindoy is supported by the Air Force Office of Scientific
Research (Air Force Materiel Command, USAF award no. FA9550-14-1-0095), a Perkin Research 
 Studentship from Magdalen College, Oxford, an Eleanor Sophia Wood Postgraduate Research Travelling Scholarship from the University of Sydney, and a James Fairfax Oxford Australia Scholarship.
 \end{acknowledgments}

\appendix
\section{An efficient implementation of RPMD for system-bath problems} \label{GLE_derivation}

In this appendix we describe an efficient implementation of RPMD and other path-integral based methods for system-bath problems with harmonic baths, which is based on analytically integrating out the effect of the bath in the ring-polymer picture. In Section~A1 we give an overview of the approach, detailing the key equations that were implemented. In Section~A2 we show how one can integrate out the effect of the bath in RPMD/Wolynes theory for the general case of an arbitrary number of system coordinates coupled to a harmonic bath. In Section~A3 we specialise the general result for the model problem considered in section~\ref{spin_b_sec}. Finally, in Section~A4, we describe how we numerically integrated the GLE satisfied by the system variables in RPMD.

\subsection{Overview} \label{overview}
Both the RPMD and Wolynes theory calculations for the spin-boson model can be transformed into trivial problems, which only explicitly include one nuclear degree of freedom, the reaction coordinate. Starting from the diabatic potentials [Eqs.~(\ref{reaction-coorda})-(\ref{reaction-coordc})], we can write both the RPMD and Wolynes theory Hamiltonians in the form
\begin{equation}
H^{(n)}(\mathbf{P},\mathbf{Q},\mathbf{p},\mathbf{q}) =  H^{(n)}_s(\mathbf{P},\mathbf{Q}) + H^{(n)}_{sb}(\mathbf{Q},\mathbf{p},\mathbf{q})
\end{equation}
where
\begin{equation}
\begin{aligned}
\!\!\!\!H^{(n)}_{sb}(\mathbf{Q},\mathbf{p},\mathbf{q})\!\! =\sum_{j=1}^n \sum_{\nu=1}^{N_b} \Bigg[& \frac{p_{j,\nu}^2}{2m_\nu} +\frac{1}{2}m_\nu\omega_n^2(q_{j+1,\nu}-q_{j,\nu})^2\!\\
&+\frac{1}{2}m_\nu\omega_\nu^2\Bigg(q_{j,\nu}- \frac{c_{\nu}Q_{j}}{m_\nu\omega_\nu^2}\Bigg)^2\Bigg].\!\!
\end{aligned}
\end{equation}
Integrating out the effect of the bath on the system allows us to write an effective Hamiltonian which only contains the reaction coordinate. For the Ohmic spectral density this renormalised ring-polymer Hamiltonian is simply
\begin{equation}
\tilde{H}_s^{(n)}(\mathbf{P},\mathbf{Q}) = H^{(n)}_s(\mathbf{P},\mathbf{Q}) + \sum_{k=0}^{n-1} \frac{1}{2}\gamma \omega_k {\tilde{Q}}_k^2 
\end{equation}
where $\tilde{Q}_k$ and $\omega_k=2\omega_n \sin(k\pi/n)$ are the position and frequency of the $k^{\mathrm{th}}$ normal mode of the free ring-polymer along the reaction coordinate. We see that friction has the effect of increasing the stiffness of the ring-polymer springs. As discussed in the text, this is consistent with the well known effect of friction in reducing tunnelling and making the system behave more classically.\cite{Leggett84}

For static quantities such as the Wolynes rate and the quantum transition state theory (QTST) part of the RPMD rate, one can simply do path integral molecular dynamics (PIMD) with the renormalised system Hamiltonian in Eq.~(A3). However, when evaluating dynamical quantities such as the RPMD transmission coefficient, the dynamical effect of the bath must also be included. This results in a Generalised Langevin Equation (GLE) for the motion along the reaction coordinate, which for the problem considered in section \ref{spin_b_sec} takes the form
\begin{subequations}
\begin{align}
\dot{\tilde{Q}}_{k} =&\tilde{P}_{k} \\
\dot{\tilde{P}}_{k} =& - \frac{\partial \tilde{H}_s^{(n)}}{\partial \tilde{Q}_{k}} -\int_0^t  {K}_{k}(t-\tau){{\tilde{P}}}_{k}(\tau)\mathrm{d}\tau+F_{k}(t)
\end{align}
\end{subequations}
where
\begin{equation}
\begin{aligned}
K_{k}(t)= &\,\,2\gamma \delta(t) - \gamma\omega_k \\
&+ \gamma \omega_k^2 t \mathrm{J}_0(\omega_k t)(1-\frac{\pi}{2} \mathrm{H}_1(\omega_k t))\\
&- \gamma\omega_k\mathrm{J}_1(\omega_k t)(1-\frac{\pi}{2} \omega_k t \mathrm{H}_0(\omega_k t)) 
\end{aligned}
\end{equation}
is the friction kernel, $\mathrm{J}_n(x)$ is the $n^{\mathrm{th}}$ Bessel function of the first kind, and $\mathrm{H}_n(x)$ is the $n^{\mathrm{th}}$ Struve H function. To integrate these equations of motion we use a modified form of the algorithm suggested by Berkowitz, Morgan and McCammon\cite{Berkowitz83} which makes use of the exact evolution of the free ring-polymer (as detailed in section~A4). Note that, since $\omega_0=0$, the centroid friction kernel is proportional to a delta function, and so the non-Markovian behaviour originates in the internal ring-polymer modes. This reflects the fact that whilst the Ohmic spectral density in the classical limit corresponds to Markovian dynamics of the reaction coordinate, this is not true quantum mechanically. 

\subsection{Derivation} \label{derivation}

Here we derive the Generalised Langevin Equation (GLE) for a ring-polymer Hamiltonian with $N_s$ physical system coordinates which are linearly coupled to $N_b$ harmonic oscillators, applying the approach of Cort\'es \emph{et al.}\cite{Cortes85} to the ring-polymer system bath model. We begin by writing the total ring-polymer Hamiltonian as
\begin{equation}
H^{(n)}(\mathbf{P},\mathbf{Q},\mathbf{p},\mathbf{q}) =  H^{(n)}_s(\mathbf{P},\mathbf{Q}) + H^{(n)}_{sb}(\mathbf{Q},\mathbf{p},\mathbf{q})
\end{equation}
where ($\mathbf{Q},\mathbf{P}$) and ($\mathbf{q},\mathbf{p}$) denote coordinates and momenta of the system and bath respectively, and
\begin{equation}
 \begin{aligned}
\!\!H^{(n)}_{sb}(\mathbf{Q},\mathbf{p},\mathbf{q}) =&\sum_{j=1}^n \sum_{\nu=1}^{N_b} \Bigg[ \frac{p_{j,\nu}^2}{2m_\nu} +\frac{1}{2}m_\nu\omega_n^2(q_{j+1,\nu}-q_{j,\nu})^2\!\!\!\! \\
&+\frac{1}{2}m_\nu\omega_\nu^2\Bigg(q_{j,\nu}-\sum_{\mu=1}^{N_s} \frac{c_{\mu\nu}Q_{j,\mu}}{m_\nu\omega_\nu^2}\Bigg)^2\Bigg].
\end{aligned}
\end{equation}
Throughout we use Latin subscripts to denote ring-polymer coordinates and Greek subscripts for physical degrees of freedom. Transforming from the bead representation to the normal mode representation of the free ring-polymer gives
\begin{equation}
\begin{aligned}
H^{(n)}_{sb}(\mathbf{\tilde{Q}},\mathbf{\tilde{q}},\mathbf{\tilde{p}}) =& \sum_{k=0}^{n-1} \sum_{\nu=1}^{N_b}\Bigg[ \frac{\tilde{p}_{k,\nu}^2}{2m_\nu} +\frac{1}{2}m_\nu\omega_k^2\tilde{q}_{k,\nu}^2\\
&+ \frac{1}{2}m_\nu\omega_\nu^2\Bigg(\tilde{q}_{k,\nu}-\sum_{\mu=1}^{N_s} \frac{c_{\mu\nu}\tilde{Q}_{k,\mu}}{m_\nu\omega_\nu^2}\Bigg)^2\Bigg],
\end{aligned}
\end{equation}
where again $\omega_k=2\omega_n \sin(k\pi/n)$ are the normal mode frequencies of the free ring-polymer. From this it is clear that the bath does not couple the normal modes of the system ring-polymer and hence it is straightforward to integrate out the effect of the bath on the system in this representation. The equations of motion are
\begin{subequations}
\begin{align}
&\dot{\tilde{Q}}_{k,\lambda} = \frac{\tilde{P}_{k,\lambda}}{m_\lambda},\\
&\dot{\tilde{P}}_{k,\lambda} = - \frac{\partial H^{(n)}_s}{\partial \tilde{Q}_{k,\lambda}} +\sum_{\nu=1}^{N_b} c_{\lambda\nu}\Bigg(\tilde{q}_{k,\nu}-\frac{\bm{c}_\nu\cdot\bm{\tilde{Q}}_k}{m_\nu\omega_\nu^2}\Bigg),\label{momentum_deriv}\\
&\dot{\tilde{q}}_{k,\nu} = \frac{\tilde{p}_{k,\nu}}{m_\nu},\\
&\dot{\tilde{p}}_{k,\nu} = - m_\nu\tilde{\omega}_{k,\nu}^2\tilde{q}_{k,\nu} + \bm{c}_\nu\cdot\bm{\tilde{Q}}_k,
\end{align}
\end{subequations}
where $\tilde{\omega}_{k,\nu}=\sqrt{\omega_\nu^2+\omega_k^2}$ and we have rewritten $\sum_{\mu=1}^{N_s} c_{\mu\nu}\tilde{Q}_{k,\mu}=\bm{c}_\nu\cdot\bm{\tilde{Q}}_k$.

The equations of motion for the bath degrees of freedom 
 \begin{equation}
 \ddot{\tilde{q}}_{k,\nu} + \tilde{\omega}_{k,\nu}^2\tilde{q}_{k,\nu} =  \frac{\bm{c}_\nu\cdot\bm{\tilde{Q}}_k}{m_\nu}, \label{differential_eqn}
\end{equation}
can be solved using the standard method of undetermined coefficients. The corresponding homogeneous differential equation is
 \begin{equation}
 \ddot{\tilde{q}}_{k,\nu} + \tilde{\omega}_{k,\nu}^2\tilde{q}_{k,\nu} =  0,
\end{equation}
giving the complementary function
 \begin{equation}
\tilde{q}_{k,\nu}(t) = a_{k,\nu}\cos(\tilde{\omega}_{k,\nu}t)+b_{k,\nu}\sin(\tilde{\omega}_{k,\nu}t).
\end{equation}
Hence we consider the particular solution
\begin{equation}
\tilde{q}_{k,\nu}(t) = a_{k,\nu}(t)\cos(\tilde{\omega}_{k,\nu}t)+b_{k,\nu}(t)\sin(\tilde{\omega}_{k,\nu}t),\label{particular_soln}
\end{equation}
and require that
\begin{equation}
\dot{a}_{k,\nu}(t)\cos(\tilde{\omega}_{k,\nu}t)+\dot{b}_{k,\nu}(t)\sin(\tilde{\omega}_{k,\nu}t)=0. \label{constraint}
\end{equation}
Differentiating Eq.~(\ref{particular_soln}) twice and using Eq.~(\ref{constraint}) gives
\begin{equation}
\begin{aligned}
\dot{\tilde{q}}_{k,\nu}(t) =& -\tilde{\omega}_{k,\nu}a_{k,\nu}(t)\sin(\tilde{\omega}_{k,\nu}t)\\
&+\tilde{\omega}_{k,\nu}b_{k,\nu}(t)\cos(\tilde{\omega}_{k,\nu}t)
\end{aligned}
\end{equation}
and then
\begin{equation}
\begin{aligned}
\ddot{\tilde{q}}_{k,\nu}(t) = &-\tilde{\omega}_{k,\nu}^2a_{k,\nu}(t)\cos(\tilde{\omega}_{k,\nu}t)\\
&-\tilde{\omega}_{k,\nu}^2b_{k,\nu}(t)\sin(\tilde{\omega}_{k,\nu}t)\\&-\tilde{\omega}_{k,\nu}\dot{a}_{k,\nu}(t)\sin(\tilde{\omega}_{k,\nu}t)\\
&+\tilde{\omega}_{k,\nu}\dot{b}_{k,\nu}(t)\cos(\tilde{\omega}_{k,\nu}t).
\end{aligned}
\end{equation}
Hence using Eq.~(\ref{differential_eqn}) it follows that
\begin{equation}
\begin{aligned}
\frac{\bm{c}_\nu\cdot\bm{\tilde{Q}}_k(t)}{m_\nu} = &-\tilde{\omega}_{k,\nu}\dot{a}_{k,\nu}(t)\sin(\tilde{\omega}_{k,\nu}t)\\
&+\tilde{\omega}_{k,\nu}\dot{b}_{k,\nu}(t)\cos(\tilde{\omega}_{k,\nu}t),
\end{aligned}
\end{equation}
which can be rearranged using Eq.~(\ref{constraint}) to give
\begin{subequations}
\begin{align}
& \frac{\bm{c}_\nu\cdot\bm{\tilde{Q}}_k(t)}{m_\nu\tilde{\omega}_{k,\nu}} \sin(\tilde{\omega}_{k,\nu}t)= -\dot{a}_{k,\nu}(t)\\
& \frac{\bm{c}_\nu\cdot\bm{\tilde{Q}}_k(t)}{m_\nu\tilde{\omega}_{k,\nu}} \cos(\tilde{\omega}_{k,\nu}t)= \dot{b}_{k,\nu}(t).
\end{align}
\end{subequations}
Integrating gives
\begin{subequations}
\begin{align}
&{a}_{k,\nu}(t)=a_{k,\nu}(0)-\int_0^t\frac{\bm{c}_\nu\cdot\bm{\tilde{Q}}_k(\tau)}{m_\nu\tilde{\omega}_{k,\nu}} \sin(\tilde{\omega}_{k,\nu}\tau)\mathrm{d}\tau\\
&{b}_{k,\nu}(t)=b_{k,\nu}(0)+\int_0^t\frac{\bm{c}_\nu\cdot\bm{\tilde{Q}}_k(\tau)}{m_\nu\tilde{\omega}_{k,\nu}} \cos(\tilde{\omega}_{k,\nu}\tau)\mathrm{d}\tau,
\end{align}
\end{subequations}
which when substituted into Eq.~(\ref{particular_soln}) gives 
\begin{equation}
\begin{aligned}
\tilde{q}_{k,\nu}(t) =& \int_0^t \frac{\bm{c}_\nu\cdot\bm{\tilde{Q}}_k(\tau)}{m_\nu\tilde{\omega}_{k,\nu}} \sin(\tilde{\omega}_{k,\nu}(t-\tau))\mathrm{d}\tau\\
&+a_{k,\nu}(0)\cos(\tilde{\omega}_{k,\nu}t)+b_{k,\nu}(0)\sin(\tilde{\omega}_{k,\nu}t).
\end{aligned}
\end{equation}
Finally, integrating by parts and inserting the appropriate boundary conditions gives
\begin{equation}
\begin{aligned}
\tilde{q}_{k,\nu}(t) =& \frac{\bm{c}_\nu\cdot\bm{\tilde{Q}}_k(t)}{m_\nu\tilde{\omega}_{k,\nu}^2}-\frac{\bm{c}_\nu\cdot\bm{\tilde{Q}}_k(0)}{m_\nu\tilde{\omega}_{k,\nu}^2} \cos(\tilde{\omega}_{k,\nu}t)\\
&-\int_0^t\frac{\bm{c}_\nu\cdot\dot{\bm{\tilde{Q}}}_k(\tau)}{m_\nu\tilde{\omega}_{k,\nu}^2} \cos(\tilde{\omega}_{k,\nu}(t-\tau))\mathrm{d}\tau\\
&+\tilde{q}_{k,\nu}(0)\cos(\tilde{\omega}_{k,\nu}t)+\frac{\tilde{p}_{k,\nu}(0)}{m_{\nu}\tilde{\omega}_{k,\nu}}\sin(\tilde{\omega}_{k,\nu}t).
\end{aligned}
\end{equation}

Having solved for $\tilde{q}_{k,\nu}(t)$, we can now substitute this back into Eq.~(\ref{momentum_deriv}) to obtain the GLE for an arbitrary system bath model,
\begin{equation}
\bm{\dot{\tilde{P}}}_{k}(t) = - \frac{\partial \tilde{H}^{(n)}_s}{\partial \bm{\tilde{Q}}_{k}} -\int_0^t  \mathbf{K}^{(k)}(t-\tau)\cdot\dot{\bm{\tilde{Q}}}_{k}(\tau)\mathrm{d}\tau+\mathbf{F}^{(k)}(t).
\end{equation}
Here we have defined
\begin{equation}
\frac{\partial \tilde{H}^{(n)}_s}{\partial \bm{\tilde{Q}}_{k}} =\frac{\partial {H}^{(n)}_s}{\partial \bm{\tilde{Q}}_{k}}+\boldsymbol{\alpha}^{(k)}\cdot\bm{\tilde{Q}}_{k}
\end{equation}
with
\begin{equation}
{\alpha}^{(k)}_{\lambda\mu}=\sum_{\nu=1}^{N_b}\Bigg(\frac{c_{\lambda\nu}{c}_{\mu\nu}}{m_\nu{\omega}_{\nu}^2}-\frac{c_{\lambda\nu}{c}_{\mu\nu}}{m_\nu\tilde{\omega}_{k,\nu}^2}\Bigg),
\end{equation}
which accounts for the renormalisation of the system Hamiltonian due to the presence of the bath, and we have collected together the remaining effects of the bath into a friction kernel
\begin{equation}
{K}^{(k)}_{\lambda\mu}(t-\tau)=\sum_{\nu=1}^{N_b}\frac{c_{\lambda\nu}{c}_{\mu\nu}}{m_\nu\tilde{\omega}_{k,\nu}^2} \cos(\tilde{\omega}_{k,\nu}(t-\tau))
\end{equation}
and a fluctuating force
\begin{equation}
\begin{aligned}
\!\!\!\!F^{(k)}_{\lambda}(t)=\sum_{\nu=1}^{N_b}c_{\lambda\nu}\Bigg(&\Big(\tilde{q}_{k,\nu}(0)-\frac{\bm{c}_\nu\cdot\bm{\tilde{Q}}_k(0)}{m_\nu\tilde{\omega}_{k,\nu}^2}\Big) \cos(\tilde{\omega}_{k,\nu}t)\\
&+\frac{\tilde{p}_{k,\nu}(0)}{m_\nu\tilde{\omega}_{k,\nu}}\sin(\tilde{\omega}_{k,\nu}t)\Bigg).
\end{aligned}
\end{equation}

It is straightforward to show that, for a system at thermal equilibrium, the fluctuating force and the friction kernel are related by the fluctuation-dissipation theorem
\begin{equation}
\langle F^{(k)}_{\lambda}(0){F}^{(k)}_{\mu}(t)\rangle = \frac{1}{\beta_n} {K}^{(k)}_{\lambda\mu}(t).
\end{equation}
The easiest way to see this is to note that the ring polymer Hamiltonian in Eq.~(A6) can be rewritten as
\begin{equation}
H^{(n)} = \tilde{H}_s^{(n)} + \tilde{H}_{sb}^{(n)}
\end{equation}
where each term is written in renormalised form as
\begin{subequations}
\begin{align}
\!\!\!\!\!\tilde{H}_s^{(n)} &= H^{(n)}_s + \sum_{k=0}^{n-1} \frac{1}{2}\tilde{\bm{Q}}_k^T\bm{\alpha}^{(k)} \tilde{\bm{Q}}_k \\
\!\!\!\!\tilde{H}_{sb}^{(n)} &= \sum_{k=0}^{n-1} \sum_{\nu=1}^{N_b}\!\!\Bigg[ \frac{\tilde{p}_{k,\nu}^2}{2m_\nu}\!\! +\!\! \frac{1}{2}m_\nu\tilde{\omega}_{k,\nu}^2\Bigg(\tilde{q}_{k,\nu}- \frac{\bm{c}_\nu\cdot\bm{\tilde{Q}}_k}{m_\nu\tilde{\omega}_{k,\nu}^2}\Bigg)^{\!\!\!\!\!2}\Bigg].
\end{align}
\end{subequations}
Since the fluctuating force in Eq.~(A26) only depends on the bath degrees of freedom, and since the result of doing so is independent of ${\bf Q}(0)\equiv \left\{\bm{\tilde{Q}}_k(0)\right\}_0^{n-1}$, the thermally averaged force-force correlation function can be evaluated with the distribution $e^{-\beta_n \tilde{H}_{sb}(t=0)}/\tr[e^{-\beta_n \tilde{H}_{sb}(t=0)}]$, which leads directly to Eq.~(A27). Note also that, in the infinite bath limit, the fluctuating force is Gaussian (by the Central Limit Theorem), and since $\bigl<F_{\lambda}^{(k)}(0)\bigr>=0$, a knowledge of the friction kernel completely specifies the form of the fluctuating force.

Finally, we note that the expressions for $\alpha_{\lambda\mu}^{(k)}$ and $K_{\lambda\mu}^{(k)}(t)$ in Eqs.~(A24) and~(A25) can be recast in terms of the spectral density
\begin{equation}
J_{\lambda\mu}(\omega)=\frac{\pi}{2}  \sum_{\nu=1}^{N_b} \frac{c_{\lambda\nu}c_{\mu\nu}}{m_\nu{\omega}_{\nu}} \delta(\omega-\omega_\nu),
\end{equation}
either as
\begin{equation}
\alpha^{(k)}_{\lambda\mu} = \frac{2}{\pi} \int_0^\infty \frac{J_{\lambda\mu}(\omega)}{\omega}- \frac{J_{\lambda\mu}(\omega)\omega}{\omega^2+\omega_k^2} \mathrm{d}\omega
\end{equation}
and 
\begin{equation}
K^{(k)}_{\lambda\mu}(t) = \frac{2}{\pi}\int_{0}^\infty \frac{J_{\lambda\mu}(\omega)\omega}{\omega^2+\omega_k^2}\cos(\sqrt{\omega^2+\omega_k^2}\, t) \mathrm{d}\omega,
\end{equation}
or equivalently as
\begin{equation}
\alpha^{(k)}_{\lambda\mu} = \frac{2}{\pi} \int_0^\infty \frac{J_{\lambda\mu}(\omega)}{\omega}-\frac{J^{(k)}_{\lambda\mu}(\omega)}{\omega}\mathrm{d}\omega
\end{equation}
and 
\begin{equation}
K^{(k)}_{\lambda\mu}(t) = \frac{2}{\pi}\int_{0}^\infty \frac{J^{(k)}_{\lambda\mu}(\omega)}{\omega}\cos(\omega t)\mathrm{d}\omega,
\end{equation}
where we have defined
\begin{equation}
J^{(k)}_{\lambda\mu}(\omega)=\theta(\omega-\omega_k)J_{\lambda\mu}(\sqrt{\omega^2-\omega_k^2})
\end{equation}
with $\theta(x)$ being the Heaviside step function.

\subsection{One Dimensional System with Ohmic Spectral Density} \label{Ohmic}

The problem considered in Section~\ref{spin_b_sec} consists of a single system coordinate (the reaction coordinate) coupled to a bath with an Ohmic spectral density $J(\omega)=\gamma \omega$. The frequencies which enter the renormalisation of the system Hamiltonian are therefore simply
\begin{equation}
\alpha_k = \frac{2}{\pi} \int_0^\infty \frac{\gamma \omega_k^2}{\omega^2 + \omega_k^2} \mathrm{d}\omega ,
\end{equation}
where, since there is only one physical system coordinate, we have changed notation slightly so that the subscript refers to the normal mode. This integral is easily evaluated to give
\begin{equation}
 \alpha_k =\gamma \omega_k.
 \end{equation}
The friction kernel can also be evaluated analytically for this problem by considering
\begin{equation}
K_k(t) = \frac{2}{\pi} \int_{0}^{\infty} \frac{J_k(\omega)}{\omega} \cos(\omega t) \mathrm{d}\omega
\end{equation}
with
\begin{equation}
J_k(\omega) = \theta(\omega-\omega_k)\gamma \sqrt{\omega^2-\omega_k^2}.
\end{equation}
This can be rearranged into the form
\begin{equation}
\begin{aligned}
K_k(t) =  & \frac{2}{\pi} \int_{\omega_k}^{\infty} \frac{\gamma \sqrt{\omega^2-\omega_k^2}-\gamma \omega}{\omega} \cos(\omega t) \mathrm{d}\omega \\
&+2\gamma\delta(t)-\frac{2}{\pi}\int_{0}^{\omega_k} \gamma \cos(\omega t) \mathrm{d}\omega,
\end{aligned}
\end{equation}
and then straightforwardly evaluated to give
\begin{equation}
\begin{aligned}
K_{k}(t)=& 2\gamma \delta(t) - \gamma\omega_k \\
& + \gamma \omega_k^2 t \mathrm{J}_0(\omega_k t)\Big(1-\frac{\pi}{2} \mathrm{H}_1(\omega_k t)\Big)\\
& - \gamma\omega_k\mathrm{J}_1(\omega_k t)\Big(1-\frac{\pi}{2} \omega_k t \mathrm{H}_0(\omega_k t)\Big) ,
\end{aligned}
\end{equation}
where $\mathrm{J}_n(x)$ is the $n^{\mathrm{th}}$ Bessel function of the first kind and $\mathrm{H}_n(x)$ is the $n^{\mathrm{th}}$ Struve H function.

The GLE for each internal mode of the ring polymer in a one-dimensional system coupled to a bath with Ohmic spectral density can thus be written as
\begin{subequations}
\begin{align}
\dot{\tilde{Q}}_{k} =&\frac{\tilde{P}_{k}}{m} \\
\dot{\tilde{P}}_{k} =&- \frac{\partial \tilde{H}_s^{(n)}}{\partial \tilde{Q}_{k}} -\frac{1}{m}\int_0^t  {K}_{k}(t-\tau){{\tilde{P}}}_{k}(\tau)\mathrm{d}\tau+F_{k}(t),
\end{align}
\end{subequations}
where
\begin{equation}
\frac{\partial \tilde{H}_s^{(n)}}{\partial \tilde{Q}_{k}} = \frac{\partial {V}_s^{(n)}}{\partial \tilde{Q}_{k}} + \omega_k^2 \tilde{Q}_{k} + \gamma \omega_k \tilde{Q}_{k}.
\end{equation}

\subsection{Generalised Langevin Equation: Numerical Integration}\label{GLE_integration}
The standard integration scheme used to evolve RPMD can be written in the form
\begin{subequations}
\begin{align}
\tilde{P}_k' & \leftarrow \tilde{P}_k^{(i)} + \frac{\Delta t}{2}f_k^{(i)}  \\
\tilde{Q}_k^{(i+1)} & \leftarrow \cos(\omega_k \Delta t) \tilde{Q}_k^{(i)} + \frac{1}{m\omega_k} \sin(\omega_k \Delta t) \tilde{P}_k' \\
\tilde{P}_k'& \leftarrow \cos(\omega_k \Delta t) \tilde{P}_k' - {m\omega_k}\sin(\omega_k \Delta t) \tilde{Q}_k^{(i)} \\
\tilde{P}_k^{(i+1)} & \leftarrow \tilde{P}_k' + \frac{\Delta t}{2}f_k^{(i+1)}
\end{align}
\end{subequations}
where $f_k^{(i)}$ is the external force on the the $k^{\mathrm{th}}$ normal mode (excluding the ring polymer spring force) at time step $i$, and $\tilde{P}_k'$ is a temporary value of the momentum (between the two time steps).
For simplicity we shall consider just a single system coordinate; the extension to more degrees of freedom is straightforward but notationally cumbersome. Following Berkowitz \emph{et al.},\cite{Berkowitz83} we start by considering the standard ring-polymer integration scheme in its position only form (analogous to the difference between the Verlet and velocity Verlet algorithms)
\begin{subequations}
\begin{align}
\tilde{Q}_k^{(i+1)} &\leftarrow 2  {C}_k \tilde{Q}_k^{(i)} - \tilde{Q}_k^{(i-1)} + f_k^{(i)} \frac{\Delta t^2}{m} {S}_k \\
\tilde{P}_k^{(i)} &\leftarrow \frac{m}{{S}_k}\frac{ \tilde{Q}_k^{(i+1)}-\tilde{Q}_k^{(i-1)} }{2\Delta t } \label{verlet_mom},
\end{align}
\end{subequations}
where ${C}_k=\cos(\omega_k \Delta t)$ and ${S}_k=\frac{1}{\omega_k\Delta t}\sin(\omega_k \Delta t)$. 

Firstly we incorporate the renormalisation of the Hamiltonian in the exact ring-polymer evolution by replacing $C_k\to\tilde{C}_k=\cos(\tilde{\omega}_k\Delta t)$ and $S_k\to\tilde{S}_k=\frac{1}{\tilde{\omega}_k\Delta t}\sin(\tilde{\omega}_k\Delta t)$, with $\tilde{\omega}_k = \sqrt{\omega_k^2+\gamma\omega_k}$. Then we use the fact that the external force 
\begin{equation}
f_k(t) = - \frac{\partial {V}_s^{(n)}}{\partial \tilde{Q}_{k}} -\frac{1}{m}\int_0^t  {K}_{k}(t-\tau){{\tilde{P}}}_{k}(\tau)\mathrm{d}\tau + F_k(t) 
\end{equation}
can be discretised using the trapezium rule to give
\begin{equation}
f_k^{(i)} \simeq - \frac{\partial {V}_s^{(n)}}{\partial \tilde{Q}^{(i)}_{k}} - \frac{1}{m}\sum_{j=0}^{i} w_j  {K}_{k}^{(j)}{{\tilde{P}}}_{k}^{(i-j)}\Delta t  + F_k^{(i)}, 
\end{equation}
 where $w_j$ are the integration weights and $F_k^{(i)}$ is a particular realisation of the fluctuating force, the generation of which is discussed later. Using this with Eq.~(\ref{verlet_mom}) for the momenta gives
\begin{equation}
\begin{aligned}
\!\!\!\tilde{Q}_k^{(i+1)}(1+\frac{\Delta t^2}{4m}K^{(0)}_{k}) = &2  \tilde{C}_k \tilde{Q}_k^{(i)} - \tilde{Q}_k^{(i-1)}(1-\frac{\Delta t^2}{4m}K^{(0)}_{k})\\
& - \frac{1}{m}\sum_{j=1}^{i} w_j  {K}_{k}^{(j)}{{\tilde{P}}}_{k}^{(i-j)}\Delta t  \frac{\Delta t^2}{m} \tilde{S}_k  \\
&- \frac{\partial {V}_s^{(n)}}{\partial \tilde{Q}^{(i)}_{k}} \frac{\Delta t^2}{m} \tilde{S}_k +  F_k^{(i)}  \frac{\Delta t^2}{m} \tilde{S}_k .
\end{aligned}
\end{equation}
 Finally, we rearrange this into a form analogous to the usual ring polymer integration scheme to obtain
 \begin{subequations}
\begin{align}
\tilde{P}_k' &\leftarrow \tilde{P}_k^{(i)}\Big(1-\frac{\Delta t^2}{4m}K^{(0)}_{k}\Big) \label{begining_step_1}   \\
\tilde{P}_k' &\leftarrow \tilde{P}_k' - \frac{\Delta t^2}{2m}\sum_{j=1}^{i} w_j  {K}_{k}^{(j)}{{\tilde{P}}}_{k}^{(i-j)} + \frac{\Delta t}{2}F_k^{(i)} \label{begining_step_2}  \\
\tilde{P}_k' &\leftarrow \tilde{P}_k' - \frac{\Delta t}{2}\frac{\partial {V}_s^{(n)}}{\partial \tilde{Q}^{(i)}_{k}} \label{middle_step_1}  \\
\tilde{Q}_k^{(i+1)} &\leftarrow \cos(\tilde{\omega}_k \Delta t) \tilde{Q}_k^{(i)} + \frac{1}{m\tilde{\omega}_k} \sin(\tilde{\omega}_k \Delta t) \tilde{P}_k' \\
\tilde{P}_k' &\leftarrow \cos(\tilde{\omega}_k \Delta t) \tilde{P}_k' - {m\tilde{\omega}_k}\sin(\tilde{\omega}_k \Delta t) \tilde{Q}_k^{(i)} \\
\tilde{P}_k' &\leftarrow \tilde{P}_k'- \frac{\Delta t}{2}\frac{\partial {V}_s^{(n)}}{\partial \tilde{Q}^{(i+1)}_{k}}  \label{middle_step_4} \\
\tilde{P}_k' &\leftarrow \tilde{P}_k' - \frac{\Delta t^2}{2m}\sum_{j=1}^{i+1} w_j  {K}_{k}^{(j)}{{\tilde{P}}}_{k}^{(i+1-j)} + \frac{\Delta t}{2}F_k^{(i+1)} \label{end_step_1} \\
\tilde{P}_k^{(i+1)} &\leftarrow \tilde{P}_k'\Big(1+\frac{\Delta t^2}{4m}K^{(0)}_{k}\Big)^{-1}. \label{end_step_2}
\end{align}
\end{subequations}
This can be implemented by noting that Eqs.~(\ref{middle_step_1})-(\ref{middle_step_4}) are just the usual integration scheme with modified spring constants, to which one simply needs to add the thermostatting steps in Eqs.~(\ref{begining_step_1}), (\ref{begining_step_2}), (\ref{end_step_1}) and (\ref{end_step_2}). We also note that, since the dynamics of the centroid ($k=0$) mode is Markovian, we can simply use the usual path integral Langevin equation (PILE)\cite{Ceriotti10} integration scheme for this mode with the appropriate friction constant. 

As described by Berkowitz \emph{et al.},\cite{Berkowitz83} the fluctuating force $F_k(t)$ can be obtained on the necessary time grid using a discrete Fourier transform. This can be achieved by first writing the fluctuating force as
\begin{equation}
F_k(t)\! =\! \int_{0}^{\infty}\!\! G_k(\omega)\Big(\! \xi_a(\omega)\cos(\omega t)\! +\! \xi_b(\omega)\sin(\omega t)\!\Big)\mathrm{d}\omega,\!\!\!
\end{equation}
where $\xi_a(\omega)$ and $\xi_b(\omega)$ are Gaussian random noise terms satisfying
\begin{subequations}
\begin{align}
& \langle \xi_a(\omega) \rangle = \langle \xi_b(\omega) \rangle = 0 \\
&\langle \xi_a(\omega)\xi_a(\omega')\rangle = \delta(\omega-\omega') \\
&\langle \xi_b(\omega)\xi_b(\omega')\rangle = \delta(\omega-\omega') \\
&\langle \xi_a(\omega)\xi_b(\omega')\rangle=0,
\end{align}
\end{subequations}
and 
\begin{equation}
G_k(\omega) = \sqrt{\frac{2}{\pi\beta_n} \frac{J_k(\omega)}{\omega}} .
\end{equation}
More formally $\xi_a(\omega)$ and $\xi_b(\omega)$ are derivatives of Wiener processes, $w_a(\omega)$ and $w_b(\omega)$ respectively, and as such we can write
\begin{equation}
\begin{aligned}
F_k(t) =& \int_{0}^{\infty} G_k(\omega) \cos(\omega t) \mathrm{d}w_a(\omega)\\
&+\int_{0}^{\infty} G_k(\omega) \sin(\omega t) \mathrm{d}w_b(\omega).
\end{aligned}
\end{equation} 
Here the integrals are Stratanovich stochastic integrals defined as
\begin{equation}
\begin{aligned}
\!\!\!\!\int_0^{\omega_{\rm{max}}}\!\! g(\omega)\mathrm{d}w(\omega)\! =\! \underset{{N\to\infty}}{\text{ms-lim}}\Bigg[&\sum_{j=0}^{N-1}\frac{g(\omega_j)+g(\omega_{j+1})}{2}\\
&\times\big(w(\omega_{j+1})-w(\omega_{j})\big)\Bigg],
\end{aligned}
\end{equation}
in which the nodes $\omega_j$ form an ordered subdivision $0=\omega_0<\omega_1<\cdots \omega_N=\omega_{\rm max}$ of the interval $[0,\omega_{\rm{max}}]$ and \lq$\text{ms-lim}$' denotes a limit in mean square. Note that from the definition of a Wiener process
\begin{equation}
w(\omega_{j+1})-w(\omega_j)=\sqrt{(\omega_{j+1}-\omega_j)}\xi^{(j)}
\end{equation}
where $\{\xi^{(j)};\,j=0,\dots ,N \}$ is a Gaussian stochastic process with zero mean and unit variance. 

Since we wish to obtain $F_k(t)$ on a discrete time grid, with spacing $\Delta t$, we truncate the integrals in Eq.~(A53) at the Nyquist critical frequency
\begin{equation}
\begin{aligned}
F_k(i\Delta t) \simeq &\int_0^{\frac{\pi}{\Delta t}} G_k(\omega)\cos(\omega i\Delta t)\mathrm{d}w_a(\omega)\\&+\int_{0}^{\frac{\pi}{\Delta t}} G_k(\omega) \sin(\omega i\Delta t) \mathrm{d}w_b(\omega) 
\end{aligned}
\end{equation}
and then take a finite $N$ approximation to each integral, using a evenly spaced grid in frequency space with $\omega_j=j\Delta \omega$ and $\Delta \omega = \frac{\pi}{N\Delta t}$. This gives
\begin{equation}
F_k(i\Delta t)\!\simeq\! F_k^{(i)}\! =\! \sum_{j=0}^{N} a^{(j)}_k\cos(ij\pi/N) + b^{(j)}_k\sin(ij\pi/N),\!\!
\end{equation}
where we have defined
\begin{subequations}
\begin{align}
&a_k^{(j)} = \begin{cases} G_k(\omega_j)  \xi_a^{(j)} \sqrt{\Delta\omega} &\text{if } 0<j<N \\
G_k(\omega_j) \xi_a^{(j)} \frac{\sqrt{\Delta\omega}}{2} & \text{if } j=0\text{ or }N
\end{cases}\\
&b_k^{(j)} = \begin{cases}G_k(\omega_j) \xi_b^{(j)}\sqrt{\Delta\omega} &\text{if } 0<j<N \\
G_k(\omega_j) \xi_b^{(j)} \frac{\sqrt{\Delta\omega}}{2} & \text{if } j=0\text{ or }N,
\end{cases}
\end{align}
\end{subequations}
in which $\{\xi^{(j)}_a\}$ and $\{\xi^{(j)}_b\}$ are independent sets of uncorrelated normal deviates with zero mean and unit variance.  Since this has the form of a discrete half-complex-to-real Fourier transform it is straightforward to implement numerically. This will return a series of $2N$ time points, of which only the first $N$ should be used since the average of the force-force correlation function will recur. Hence $N$ must be at least as large as the number of time-steps. Note that one must also ensure that $N$ is sufficiently large to avoid aliasing.

This scheme is very simple to implement and proved efficient enough for the present calculations. However, we note that if one were to consider a problem with a larger system dimension it might become advantageous to use a multivariate Ornstein-Uhlenbeck process, in a manner analogous to that adopted by Ceriotti {\em et al.}\cite{Ceriotti10b} in their construction of methods for solving GLEs.

\end{document}

\def\bra#1{\left<{#1}\right|}
\def\ket#1{\left|{#1}\right>}
\def\expval#1#2{\bra{#2} {#1} \ket{#2}}
\def\mapright#1{\smash{\mathop{\longrightarrow}\limits^{_{_{\phantom{X}}}{#1}_{_{\phantom{X}}}}}}

\title{On the calculation of quantum mechanical electron transfer rates:\\ Supplementary Material}
\author{Joseph E. Lawrence}
\affiliation{Department of Chemistry, University of Oxford, Physical and Theoretical Chemistry Laboratory, South Parks Road, Oxford, OX1 3QZ, UK}
\author{Theo Fletcher}
\affiliation{Department of Chemistry, University of Oxford, Physical and Theoretical Chemistry Laboratory, South Parks Road, Oxford, OX1 3QZ, UK}
\author{Lachlan P. Lindoy}
\affiliation{Department of Chemistry, University of Oxford, Physical and Theoretical Chemistry Laboratory, South Parks Road, Oxford, OX1 3QZ, UK}
\author{David E. Manolopoulos}
\affiliation{Department of Chemistry, University of Oxford, Physical and Theoretical Chemistry Laboratory, South Parks Road, Oxford, OX1 3QZ, UK}

\begin{abstract}
\end{abstract}

\maketitle

\begin{table}[H]
\center
\caption{$\beta\epsilon=0$, $\beta\Lambda=60$, $\beta\hbar\Omega=0.5$ and $\gamma=32\Omega$.}
\begin{tabular}{S|SSSS}
%$\log_{10}(k_{\mathrm{RP}})$ & $\log_{10}(k_{\mathrm{RP}})$ & $\log_{10}(k_{\mathrm{RP}})$ & $\log_{10}(k_{\mathrm{RP}})$ & $\log_{10}(k_{\mathrm{RP}})$ \\
\toprule
 &\multicolumn{4}{c}{\text{$\vphantom{\big(}\log_{10}(k\beta\hbar)$}}  \\
 
\text{$\vphantom{\big(}\log_{10}(\beta\Delta)$} & \multicolumn{1}{c}{\text{HEOM}}  & \multicolumn{1}{c}{\text{IF}} & \multicolumn{1}{c}{\text{RPMD}} & \multicolumn{1}{c}{\text{Wolynes}} \\
\midrule
-1.0   & -9.19 & -9.20  & -8.25 & -9.14 \\
-0.9 & -9.01 & -9.02 & -8.24 & -8.94 \\
-0.8 & -8.85 & -8.86 & -8.24 & -8.74 \\
-0.7 & -8.70  & -8.71 & -8.24 & -8.54 \\
-0.6 & -8.57 & -8.58 & -8.23 & -8.34 \\
-0.5 & -8.46 & -8.47 & -8.22 & -8.14 \\
-0.4 & -8.36 & -8.38 & -8.21 & -7.94 \\
-0.3 & -8.29 & -8.31 & -8.19 & -7.74 \\
-0.2 & -8.22 & -8.24 & -8.16 & -7.54 \\
-0.1 & -8.16 & -8.18 & -8.12 & -7.34 \\
0.0    & -8.10  & -8.11 & -8.07 & -7.14 \\
0.1  & -8.02 & -8.02 & -8.00    & -6.94 \\
0.2  & -7.92 & -7.92 & -7.91 & -6.74 \\
0.3  & -7.79 & -7.79 & -7.78 & -6.54 \\
0.4  & -7.62 & -7.62 & -7.62 & -6.34 \\
0.5  & -7.41 & -7.41 & -7.41 & -6.14 \\
0.6  & -7.15 & -7.15 & -7.15 & -5.94 \\
0.7  & -6.82 & -6.83 & -6.83 & -5.74 \\
0.8  & -6.42 & -6.43 & -6.43 & -5.54 \\
0.9  & -5.95 & -5.95 & -5.95 & -5.34 \\
1.0    & -5.41 & -5.39 & -5.39 & -5.14\\
 \bottomrule
\end{tabular}
\end{table}

\begin{table}[H]
\center
\caption{$\beta\epsilon=0$, $\beta\Lambda=60$, $\beta\hbar\Omega=0.5$ and $\gamma=\Omega$.}
\begin{tabular}{S|SSSS}
%$\log_{10}(k_{\mathrm{RP}})$ & $\log_{10}(k_{\mathrm{RP}})$ & $\log_{10}(k_{\mathrm{RP}})$ & $\log_{10}(k_{\mathrm{RP}})$ & $\log_{10}(k_{\mathrm{RP}})$ \\
\toprule
 &\multicolumn{4}{c}{\text{$\vphantom{\big(}\log_{10}(k\beta\hbar)$}}  \\
 
\text{$\vphantom{\big(}\log_{10}(\beta\Delta)$} & \multicolumn{1}{c}{\text{HEOM}}  & \multicolumn{1}{c}{\text{IF}} & \multicolumn{1}{c}{\text{RPMD}} & \multicolumn{1}{c}{\text{Wolynes}} \\
\midrule
-1.0   & -9.12 & -9.13 & -7.37 & -9.12 \\
-0.9 & -8.93 & -8.93 & -7.37 & -8.92 \\
-0.8 & -8.73 & -8.73 & -7.36 & -8.72 \\
-0.7 & -8.54 & -8.54 & -7.36 & -8.52 \\
-0.6 & -8.34 & -8.35 & -7.35 & -8.32 \\
-0.5 & -8.16 & -8.16 & -7.34 & -8.12 \\
-0.4 & -7.97 & -7.98 & -7.32 & -7.92 \\
-0.3 & -7.80 & -7.80 & -7.29 & -7.72 \\
-0.2 & -7.63 & -7.64 & -7.25 & -7.52 \\
-0.1 & -7.47 & -7.48 & -7.21 & -7.32 \\
0.0    & -7.32 & -7.33 & -7.14 & -7.12 \\
0.1  & -7.17 & -7.18 & -7.05 & -6.92 \\
0.2  & -7.01 & -7.02 & -6.93 & -6.72 \\
0.3  & -6.84 & -6.84 & -6.78 & -6.52 \\
0.4  & -6.63 & -6.63 & -6.59 & -6.32 \\
0.5  & -6.38 & -6.38 & -6.36 & -6.12 \\
0.6  & -6.08 & -6.08 & -6.06 & -5.92 \\
0.7  & -5.71 & -5.71 & -5.70 & -5.72 \\
0.8  & -5.27 & -5.27 & -5.27 & -5.52 \\
0.9  & -4.76 & -4.76 & -4.76 & -5.32 \\
1.0    & -4.17 & -4.17 & -4.17 & -5.12 \\
 \bottomrule
\end{tabular}
\end{table}

\begin{table}[H]
\center
\caption{$\beta\epsilon=0$, $\beta\Lambda=60$, $\beta\hbar\Omega=4$ and $\gamma=32\Omega$.}
\begin{tabular}{S|SSSS}
%$\log_{10}(k_{\mathrm{RP}})$ & $\log_{10}(k_{\mathrm{RP}})$ & $\log_{10}(k_{\mathrm{RP}})$ & $\log_{10}(k_{\mathrm{RP}})$ & $\log_{10}(k_{\mathrm{RP}})$ \\
\toprule
 &\multicolumn{4}{c}{\text{$\vphantom{\big(}\log_{10}(k\beta\hbar)$}}  \\
 
\text{$\vphantom{\big(}\log_{10}(\beta\Delta)$} & \multicolumn{1}{c}{\text{HEOM}}  & \multicolumn{1}{c}{\text{IF}} & \multicolumn{1}{c}{\text{RPMD}} & \multicolumn{1}{c}{\text{Wolynes}} \\
\midrule 
-1.0   & -9.05 & -9.05 & -7.04 & -9.05 \\
-0.9 & -8.85 & -8.85 & -7.04 & -8.85 \\
-0.8 & -8.66 & -8.65 & -7.04 & -8.65 \\
-0.7 & -8.46 & -8.46 & -7.04 & -8.45 \\
-0.6 & -8.27 & -8.26 & -7.03 & -8.25 \\
-0.5 & -8.08 & -8.07 & -7.02 & -8.05 \\
-0.4 & -7.91 & -7.88 & -7.02 & -7.85 \\
-0.3 & -7.74 & -7.70 & -7.00 & -7.65 \\
-0.2 & -7.58 & -7.53 & -6.98 & -7.45 \\
-0.1 & -7.43 & -7.37 & -6.95 & -7.25 \\
0.0    & -7.29 & -7.22 & -6.91 & -7.05 \\
0.1  & -7.16 & -7.07 & -6.86 & -6.85 \\
0.2  & -7.02 & -6.93 & -6.78 & -6.65 \\
0.3  & -6.87 & -6.77 & -6.68 & -6.45 \\
0.4  & -6.69 & -6.60 & -6.54 & -6.25 \\
0.5  & -6.46 & -6.39 & -6.35 & -6.05 \\
0.6  & -6.19 & -6.14 & -6.11 & -5.85 \\
0.7  & -5.86 & -5.82 & -5.80 & -5.65 \\
0.8  & -5.47 & -5.43 & -5.42 & -5.45 \\
0.9  & -5.00 & -4.96 & -4.95 & -5.25 \\
1.0    & -4.46 & -4.41 & -4.41 & -5.05 \\
 \bottomrule
\end{tabular}
\end{table}

\begin{table}[H]
\center
\caption{$\beta\epsilon=0$, $\beta\Lambda=60$, $\beta\hbar\Omega=4$ and $\gamma=\Omega$.}
\begin{tabular}{S|SSSS}
%$\log_{10}(k_{\mathrm{RP}})$ & $\log_{10}(k_{\mathrm{RP}})$ & $\log_{10}(k_{\mathrm{RP}})$ & $\log_{10}(k_{\mathrm{RP}})$ & $\log_{10}(k_{\mathrm{RP}})$ \\
\toprule
 &\multicolumn{4}{c}{\text{$\vphantom{\big(}\log_{10}(k\beta\hbar)$}}  \\
 
\text{$\vphantom{\big(}\log_{10}(\beta\Delta)$} & \multicolumn{1}{c}{\text{HEOM}}  & \multicolumn{1}{c}{\text{IF}} & \multicolumn{1}{c}{\text{RPMD}} & \multicolumn{1}{c}{\text{Wolynes}} \\
\midrule
 -1.0   & -8.00 & -8.00 & -4.78 & -8.00 \\
-0.9 & -7.80 & -7.80 & -4.78 & -7.80 \\
-0.8 & -7.60 & -7.60 & -4.78 & -7.60 \\
-0.7 & -7.40 & -7.40 & -4.78 & -7.40 \\
-0.6 & -7.20 & -7.20 & -4.78 & -7.20 \\
-0.5 & -7.00 & -7.00 & -4.78 & -7.00 \\
-0.4 & -6.80 & -6.80 & -4.77 & -6.80 \\
-0.3 & -6.59 & -6.59 & -4.77 & -6.60 \\
-0.2 & -6.39 & -6.39 & -4.76 & -6.40 \\
-0.1 & -6.19 & -6.18 & -4.74 & -6.20 \\
0.0    & -5.98 & -5.97 & -4.72 & -6.00 \\
0.1  & -5.77 & -5.75 & -4.69 & -5.80 \\
0.2  & -5.56 & -5.53 & -4.65 & -5.60 \\
0.3  & -5.34 & -5.30 & -4.58 & -5.40 \\
0.4  & -5.10 & -5.05 & -4.49 & -5.20 \\
0.5  & -4.84 & -4.79 & -4.37 & -5.00 \\
0.6  & -4.56 & -4.51 & -4.20 & -4.80 \\
0.7  & -4.23 & -4.18 & -3.96 & -4.60 \\
0.8  & -3.85 & -3.81 & -3.66 & -4.40 \\
0.9  & -3.41 & -3.37 & -3.27 & -4.20 \\
1.0    & -2.89 & -2.87 & -2.80 & -4.00 \\
 \bottomrule
\end{tabular}
\end{table}

\begin{table}[H]
\center
\caption{$\beta\epsilon=15$, $\beta\Lambda=60$, $\beta\hbar\Omega=0.5$ and $\gamma=32\Omega$.}
\begin{tabular}{S|SSSS}
%$\log_{10}(k_{\mathrm{RP}})$ & $\log_{10}(k_{\mathrm{RP}})$ & $\log_{10}(k_{\mathrm{RP}})$ & $\log_{10}(k_{\mathrm{RP}})$ & $\log_{10}(k_{\mathrm{RP}})$ \\
\toprule
 &\multicolumn{4}{c}{\text{$\vphantom{\big(}\log_{10}(k\beta\hbar)$}}  \\
 
\text{$\vphantom{\big(}\log_{10}(\beta\Delta)$} & \multicolumn{1}{c}{\text{HEOM}}  & \multicolumn{1}{c}{\text{IF}} & \multicolumn{1}{c}{\text{RPMD}} & \multicolumn{1}{c}{\text{Wolynes}} \\
\midrule
-1.0   & -6.34 & -6.35 & -5.44 & -6.30 \\
-0.9 & -6.17 & -6.18 & -5.44 & -6.10 \\
-0.8 & -6.01 & -6.02 & -5.44 & -5.90 \\
-0.7 & -5.86 & -5.88 & -5.43 & -5.70 \\
-0.6 & -5.74 & -5.76 & -5.43 & -5.50 \\
-0.5 & -5.63 & -5.65 & -5.42 & -5.30 \\
-0.4 & -5.54 & -5.57 & -5.40 & -5.10 \\
-0.3 & -5.47 & -5.49 & -5.39 & -4.90 \\
-0.2 & -5.41 & -5.43 & -5.36 & -4.70 \\
-0.1 & -5.36 & -5.37 & -5.32 & -4.50 \\
0.0    & -5.30 & -5.30 & -5.27 & -4.30 \\
0.1  & -5.22 & -5.23 & -5.21 & -4.10 \\
0.2  & -5.13 & -5.13 & -5.12 & -3.90 \\
0.3  & -5.02 & -5.01 & -5.00 & -3.70 \\
0.4  & -4.87 & -4.86 & -4.85 & -3.50 \\
0.5  & -4.68 & -4.67 & -4.66 & -3.30 \\
0.6  & -4.46 & -4.43 & -4.43 & -3.10 \\
0.7  & -4.18 & -4.18 & -4.18 & -2.90 \\
0.8  & -3.87 & -3.87 & -3.86 & -2.70 \\
0.9  & -3.53 & -3.51 & -3.51 & -2.50 \\
1.0    & -3.18 & -3.15 & -3.15 & -2.30 \\
 \bottomrule
\end{tabular}
\end{table}

\begin{table}[H]
\center
\caption{$\beta\epsilon=15$, $\beta\Lambda=60$, $\beta\hbar\Omega=0.5$ and $\gamma=\Omega$.}
\begin{tabular}{S|SSSS}
%$\log_{10}(k_{\mathrm{RP}})$ & $\log_{10}(k_{\mathrm{RP}})$ & $\log_{10}(k_{\mathrm{RP}})$ & $\log_{10}(k_{\mathrm{RP}})$ & $\log_{10}(k_{\mathrm{RP}})$ \\
\toprule
 &\multicolumn{4}{c}{\text{$\vphantom{\big(}\log_{10}(k\beta\hbar)$}}  \\
 
\text{$\vphantom{\big(}\log_{10}(\beta\Delta)$} & \multicolumn{1}{c}{\text{HEOM}}  & \multicolumn{1}{c}{\text{IF}} & \multicolumn{1}{c}{\text{RPMD}} & \multicolumn{1}{c}{\text{Wolynes}} \\
\midrule
-1.0   & -6.28 & -6.28 & -4.55 & -6.28 \\
-0.9 & -6.08 & -6.09 & -4.54 & -6.08 \\
-0.8 & -5.89 & -5.89 & -4.54 & -5.88 \\
-0.7 & -5.69 & -5.69 & -4.53 & -5.68 \\
-0.6 & -5.50 & -5.50 & -4.52 & -5.48 \\
-0.5 & -5.31 & -5.31 & -4.51 & -5.28 \\
-0.4 & -5.13 & -5.13 & -4.49 & -5.08 \\
-0.3 & -4.96 & -4.96 & -4.46 & -4.88 \\
-0.2 & -4.79 & -4.80 & -4.43 & -4.68 \\
-0.1 & -4.64 & -4.65 & -4.38 & -4.48 \\
0.0    & -4.49 & -4.50 & -4.32 & -4.28 \\
0.1  & -4.35 & -4.36 & -4.23 & -4.08 \\
0.2  & -4.20 & -4.20 & -4.12 & -3.88 \\
0.3  & -4.03 & -4.04 & -3.98 & -3.68 \\
0.4  & -3.83 & -3.84 & -3.81 & -3.48 \\
0.5  & -3.59 & -3.62 & -3.59 & -3.28 \\
0.6  & -3.33 & -3.35 & -3.33 & -3.08 \\
0.7  & -3.02 & -3.03 & -3.02 & -2.88 \\
0.8  & -2.67 & -2.67 & -2.67 & -2.68 \\
0.9  & -2.28 & -2.27 & -2.27 & -2.48 \\
1.0    & -1.88 & -1.86 & -1.86 & -2.28 \\
 \bottomrule
\end{tabular}
\vspace{3.35cm}
\end{table}

\begin{table}[H]
\center
\caption{$\beta\epsilon=15$, $\beta\Lambda=60$, $\beta\hbar\Omega=4$ and $\gamma=32\Omega$.}
\begin{tabular}{S|SSSS}
%$\log_{10}(k_{\mathrm{RP}})$ & $\log_{10}(k_{\mathrm{RP}})$ & $\log_{10}(k_{\mathrm{RP}})$ & $\log_{10}(k_{\mathrm{RP}})$ & $\log_{10}(k_{\mathrm{RP}})$ \\
\toprule
 &\multicolumn{4}{c}{\text{$\vphantom{\big(}\log_{10}(k\beta\hbar)$}}  \\
 
\text{$\vphantom{\big(}\log_{10}(\beta\Delta)$} & \multicolumn{1}{c}{\text{HEOM}}  & \multicolumn{1}{c}{\text{IF}} & \multicolumn{1}{c}{\text{RPMD}} & \multicolumn{1}{c}{\text{Wolynes}} \\
\midrule
-1.0   & -6.21 & -6.21 & -4.25 & -6.21 \\
-0.9 & -6.02 & -6.01 & -4.25 & -6.01 \\
-0.8 & -5.82 & -5.82 & -4.25 & -5.81 \\
-0.7 & -5.63 & -5.62 & -4.25 & -5.61 \\
-0.6 & -5.44 & -5.43 & -4.24 & -5.41 \\
-0.5 & -5.25 & -5.24 & -4.24 & -5.21 \\
-0.4 & -5.08 & -5.05 & -4.23 & -5.01 \\
-0.3 & -4.91 & -4.88 & -4.22 & -4.81 \\
-0.2 & -4.75 & -4.71 & -4.20 & -4.61 \\
-0.1 & -4.61 & -4.55 & -4.17 & -4.41 \\
0.0    & -4.48 & -4.41 & -4.13 & -4.21 \\
0.1  & -4.36 & -4.28 & -4.08 & -4.01 \\
0.2  & -4.24 & -4.14 & -4.01 & -3.81 \\
0.3  & -4.10 & -4.00 & -3.91 & -3.61 \\
0.4  & -3.94 & -3.84 & -3.78 & -3.41 \\
0.5  & -3.75 & -3.65 & -3.62 & -3.21 \\
0.6  & -3.53 & -3.44 & -3.42 & -3.01 \\
0.7  & -3.26 & -3.19 & -3.17 & -2.81 \\
0.8  & -2.95 & -2.89 & -2.88 & -2.61 \\
0.9  & -2.61 & -2.56 & -2.56 & -2.41 \\
1.0    & -2.25 & -2.24 & -2.23 & -2.21 \\
 \bottomrule
\end{tabular}

\end{table}

\begin{table}[H]
\center
\caption{$\beta\epsilon=15$, $\beta\Lambda=60$, $\beta\hbar\Omega=4$ and $\gamma=\Omega$.}
\begin{tabular}{S|SSSS}
%$\log_{10}(k_{\mathrm{RP}})$ & $\log_{10}(k_{\mathrm{RP}})$ & $\log_{10}(k_{\mathrm{RP}})$ & $\log_{10}(k_{\mathrm{RP}})$ & $\log_{10}(k_{\mathrm{RP}})$ \\
\toprule
 &\multicolumn{4}{c}{\text{$\vphantom{\big(}\log_{10}(k\beta\hbar)$}}  \\
 
\text{$\vphantom{\big(}\log_{10}(\beta\Delta)$} & \multicolumn{1}{c}{\text{HEOM}}  & \multicolumn{1}{c}{\text{IF}} & \multicolumn{1}{c}{\text{RPMD}} & \multicolumn{1}{c}{\text{Wolynes}} \\
\midrule
 -1.0   & -5.34 & -5.34 & -2.13 & -5.34 \\
-0.9 & -5.14 & -5.14 & -2.13 & -5.14 \\
-0.8 & -4.94 & -4.94 & -2.12 & -4.94 \\
-0.7 & -4.74 & -4.74 & -2.12 & -4.74 \\
-0.6 & -4.54 & -4.54 & -2.12 & -4.54 \\
-0.5 & -4.34 & -4.33 & -2.12 & -4.34 \\
-0.4 & -4.14 & -4.13 & -2.11 & -4.14 \\
-0.3 & -3.94 & -3.93 & -2.11 & -3.94 \\
-0.2 & -3.73 & -3.72 & -2.10 & -3.74 \\
-0.1 & -3.53 & -3.51 & -2.09 & -3.54 \\
0.0    & -3.33 & -3.31 & -2.07 & -3.34 \\
0.1  & -3.12 & -3.09 & -2.04 & -3.14 \\
0.2  & -2.91 & -2.88 & -2.00 & -2.94 \\
0.3  & -2.70 & -2.65 & -1.95 & -2.74 \\
0.4  & -2.47 & -2.42 & -1.87 & -2.54 \\
0.5  & -2.24 & -2.18 & -1.76 & -2.34 \\
0.6  & -1.98 & -1.93 & -1.62 & -2.14 \\
0.7  & -1.70 & -1.65 & -1.44 & -1.94 \\
0.8  & -1.40 & -1.36 & -1.21 & -1.74 \\
0.9  & -1.07 & -1.04 & -0.94 & -1.54 \\
1.0    & -0.73 & -0.75 & -0.68 & -1.34 \\
 \bottomrule
\end{tabular}
\end{table}

All rates converged to within $\sigma(\log_{10}(k\beta\hbar))=\pm0.01$.

\clearpage